\documentclass[pdflatex,sn-mathphys-num]{sn-jnl}

\usepackage{graphicx}%
\usepackage{multirow}%
\usepackage{amsmath,amssymb}%
\usepackage{amsthm}%
\usepackage{bbm}

\usepackage{comment}
\usepackage{tabularx}

\usepackage{breqn}

\usepackage{color,xcolor}

\theoremstyle{thmstyleone}%
\theoremstyle{thmstyletwo}%

\theoremstyle{thmstylethree}%

\usepackage{makecell}

\newcommand{\indi}[1]{\scalebox{1.4}{$\mathbbm{1}$}_{#1}} 

\begin{document}


\title[Direct and Indirect Influence in Social Media]{Direct and Indirect Influence on Likes in Social Media\footnote{Ivan Kozitsin: ORCID 0000-0002-2638-0131; Anton V. Proskurnikov: ORCID 0000-0001-7065-7698}}

\author[1,2,3]{\fnm{Ivan} \sur{Kozitsin}}\email{kozitsin.ivan@mail.ru}
\author*[4]{\fnm{Anton V.} \sur{Proskurnikov}}\email{anton.proskurnikov@polito.it}

\affil[1]{\orgname{V. A. Trapeznikov Institute of Control Sciences of Russian Academy of Sciences}, \orgaddress{\street{65 Profsoyuznaya street}, \city{Moscow}, \postcode{117997}, \country{Russia}}}

\affil[2]{\orgname{Moscow Institute of Physics and Technology}, \orgaddress{\street{9 Institutskiy per.}, \city{Dolgoprudny}, \postcode{141701}, \country{Russia}}}

\affil[3]{\orgname{Tomsk State University}, \orgaddress{\street{Lenina st. 36}, \city{Tomsk}, \postcode{634050}, \country{Russia}}}

\affil*[4]{\orgname{Politecnico di Torino}, \orgaddress{\city{Turin}, \postcode{10129}, \country{Italy}}}

\abstract{
The present study investigates direct and indirect social contagion mechanisms in an online social network environment. Using a large-scale dataset comprising approximately $290{,}000$ users from the VKontakte platform, we examine the factors associated with the probability that a user likes a post. Our analysis shows that, while demographic and structural characteristics of individual nodes, such as gender and degree, contribute to the observed dynamics, the strongest associations arise from activity in the user's local network. In particular, active nodes (users who have already liked the post) at distances $d = 1$ and $d = 2$ play a central role in shaping liking behavior. We find a substantial association between second-order activity and liking probability, which persists even in the absence of active direct neighbors and is consistent with indirect influence pathways in the network. No significant association is detected for nodes at  distance three or beyond. The results also support the structural diversity hypothesis: the number of connected components among active friends is a significant predictor of liking.

\vskip1mm

To quantify $d=2$ effects, we consider two second-order metrics: (i) the induced index $m$, defined as the maximal number of active friends among the \emph{active} neighbors of a focal node, as introduced by Xie et al.~(2022), and (ii) the extended induced index $m^*$, defined analogously but computed over all neighbors of the focal node. Our findings are consistent with induced percolation mechanisms, with the extended induced index exhibiting superior explanatory power.

\vskip1mm

Finally, we find that, after accounting for structural diversity and second-order activity, the number of active direct neighbors is
negatively associated with the probability of liking, suggesting an interaction between network structure, platform-mediated visibility
mechanisms, and users' limited attention.
}


\maketitle

\section{Introduction} \label{Intro}

Online social networks have transformed the information landscape, increasing societal connectivity and creating new channels for the propagation of news, rumors, opinions, and behaviors. With billions of users worldwide, social media platforms have become major venues where individuals discover content, form attitudes, and make decisions. Understanding the mechanisms underlying the spread of information and behaviors in such environments is crucial for public health campaigns, political communication, marketing, and the mitigation of misinformation. The abundance of social media data provides an important resource for computational social science, allowing researchers to investigate social contagion~\cite{centola2007complex,kramer2014} -- the process by which behaviors, emotions, and ideas spread through social connections.

Deferring a detailed literature review to the following section, we emphasize two key features of contagion in online social media. First, social contagion is often \emph{complex} rather than simple: unlike processes in which a single contact may be sufficient for transmission, adoption of behaviors, ideas, or emotions may require reinforcement from a critical mass of peers. Complex contagion is therefore inherently a higher-order phenomenon~\cite{fang2024,iacopini2019simplicial}, in which the effect of multiple peers cannot generally be decomposed into a sum of independent dyadic influences.

Second, social media platforms are not passive communication tools. Through recommendation algorithms, news feeds, and content ranking~\cite{guess2023,hodas2014simple,bakshy2015exposure}, platforms shape what users see, creating channels for \emph{multi-hop} influence that extend beyond direct social ties and may amplify complex contagion dynamics. Indirect influence in social networks has attracted increasing attention recently, yet empirical evidence has primarily come from settings other than online social media, including studies of health behaviors~\cite{christakis2007spread,christakis2008collective}, scientific collaboration networks~\cite{xie2022indirect}, and mobile carrier networks~\cite{zhang2018}.

Explicit measurements of indirect influence -- such as the effect of friends-of-friends -- in large-scale online social networks remain scarce, although several studies have noted the possibility of multi-hop pathways~\cite{Bakshy2012,ugander2012structural}. A further challenge in studying social contagion is distinguishing social influence from homophily, i.e., the tendency of similar individuals to connect. Correlations in behavior among connected individuals may arise because people influence their friends, because similar people form ties with one another, or both~\cite{aral2009pnas}. These open questions motivate the present study.

 We investigate direct and indirect influence on liking behavior in VKontakte, a major Russian social networking platform. Specifically, we ask whether the probability of liking a post depends not only on the number of active direct friends, but also on the activity of friends-of-friends, and if so, whether this association is consistent with second-degree influence or is merely mediated through direct ties. Using data from over $290{,}000$ users and over $19{,}000$ likes on $308$ local news posts, we find a strong association between liking probability and both active direct friends and active friends-of-friends, with the latter persisting even when conditioning on direct friend activity. We introduce the \emph{extended induced index} $m^*$, defined as the maximal number of active friends across \emph{all} neighbors of the focal node, which emerges as the strongest single predictor of liking behavior, outperforming both raw counts of $d=2$ active nodes and the induced index $m$ proposed by Xie et al.~\cite{xie2022indirect}.

We conjecture that the observed second-order association may be driven, at least in part, by platform-mediated content visibility mechanisms, such as recommendation and ranking algorithms, which can expose users to content endorsed by friends-of-friends. Under this interpretation, activity at distance $d=2$ increases the likelihood of exposure even in the absence of direct signals from immediate neighbors. Overall, our findings contribute to the empirical study of indirect influence in social networks and suggest that higher-order neighborhood structure plays an important role in shaping behavioral diffusion on online platforms.

The remainder of the paper is organized as follows. Section~\ref{Sec:Literature} reviews related work on complex contagion and indirect influence in social networks.  Section~\ref{Methods} describes the empirical context, data, and the network metrics used in the analysis. We present our main results in Section~\ref{Sec:Results}, followed by a discussion in Section~\ref{Sec:discuss}.

\section{Literature} \label{Sec:Literature}

In this section, we briefly review studies on how the structure of social networks shapes contagion processes. Large-scale online platforms have enabled observational studies of unprecedented scale, although such studies are subject to confounding factors such as homophily and platform-mediated exposure. The insights obtained from observational data have been supported by controlled \textit{in vitro} and \textit{in vivo} experiments.

\subsection{Complex contagions on social networks}

We begin with early observational studies of behavioral adoption in online networks. Backstrom et al.~\cite{backstrom2006group} analyzed membership dynamics in online communities and collaboration networks, showing that the adoption probability first increases with the number of adopting neighbors, then saturates and may even decline. The authors also observed that adoption dynamics depend on the local structure of the network: less clustered communities exhibit slower growth than highly clustered ones. This finding was later addressed by Kairam et al.~\cite{kairam2012life}, who distinguished between diffusion growth, driven by nodes adjacent to existing members, and non-diffusion growth, arising from nodes outside the group's local neighborhood. While clustering facilitates diffusion growth, overall group expansion is largely driven by non-diffusion mechanisms.

In parallel, controlled experimental work has provided direct evidence for reinforcement effects in social contagion.  Centola~\cite{centola2010spread} conducted a randomized experiment on a purpose-built online platform, demonstrating that the probability of adopting a health-related behavior increases with the number of adopting neighbors. Clustered networks were found to facilitate diffusion, supporting the idea that repeated exposure from several peers plays a central role in behavioral spread. These findings provide experimental validation for the distinction between simple and complex contagion, where the latter requires multiple reinforcing exposures~\cite{centola2007complex}.

Further evidence on the role of network structure was provided by Ugander et al.~\cite{ugander2012structural}, who studied how individuals decide to join Facebook in response to invitations from their existing contacts. While adoption probability increases with the number of active neighbors (those already registered on the platform), the authors showed that it depends more strongly on the number of connected components among those neighbors than on their total count. They interpreted these components as distinct social circles, arguing that individuals are more likely to adopt a behavior when they receive signals from multiple independent circles. This phenomenon, termed \emph{structural diversity}, highlights the importance of reinforcement from socially diverse sources. Aral and Nicolaides~\cite{aral2017exercise} examined the spread of running behavior in a global online network and found that structural diversity provides a strong explanation of adoption patterns, alongside embeddedness effects
whereby shared common friends facilitate adoption. They also identified gender asymmetries in peer influence: women primarily influence women, whereas men can be influenced by both men and women.

While observational studies offer large-scale evidence, they are often subject to confounding factors such as homophily and exposure bias. To address these limitations, M{\o}nsted et al.~\cite{monsted2017evidence} conducted a controlled \textit{in vivo} experiment on Twitter using coordinated bot interventions. They compared models of simple and complex contagion and found that adoption dynamics are better explained by reinforcement-based mechanisms of complex contagion. This experimental design addresses a key concern: because each additional active neighbor increases the chance of content exposure, observed reinforcement effects could in principle reflect attention dynamics rather than peer influence. Controlled bot exposures allow these mechanisms to be disentangled. More recently, Lee et al.~\cite{lee2025complex} reported causal evidence from a country-scale randomized field experiment, showing that exposure to multiple active peers leads to adoption rates exceeding those predicted by independent cascade models. They also found that social network embeddedness moderates the effectiveness of social reinforcement.

At the same time, not all forms of information diffusion follow complex contagion dynamics. Romero et al.~\cite{romero2011differences} demonstrated that different types of content exhibit distinct spreading patterns: while political hashtags require reinforcement from multiple sources, more ordinary content often spreads through single exposures, resembling simple contagion. More broadly, prior work suggests that the nature of contagion depends on the cost, novelty, and social relevance of the behavior being adopted~\cite{aral2017exercise,romero2011differences}. These differences are also reflected in the structural properties of cascades: simple contagion processes tend to produce shorter and less persistent diffusion patterns, whereas complex contagion can generate longer propagation chains when reinforcement occurs within clustered regions of the network~\cite{wang2019anomalous,centola2019influential}. This distinction also has practical implications for network interventions: while simple contagion models emphasize highly connected hubs, complex contagion processes depend more strongly on clustered structures that provide repeated signals from multiple peers. Accordingly, recent work has proposed centrality and path-length measures tailored to complex contagion~\cite{guilbeault2021topological}, and field experiments have demonstrated the effectiveness of network-based intervention strategies in rural communities~\cite{airoldi2024induction}.

Recent extensions generalize social contagion models to higher-order networks, where adoption depends not only on pairwise contacts but also on group-level interactions among multiple nodes~\cite{li2024social}. This perspective is closely related to complex contagion, since reinforcement effects naturally arise from group-level rather than purely dyadic interactions. Related models have also considered interacting contagion processes, such as the coupling between epidemic spreading and vaccine adoption under complex social contagion~\cite{de2025interplay}.
In addition to these theoretical extensions, recent work has examined how platform design shapes contagion dynamics. Komander et al.~\cite{komander2025modeling} model link recommendation as a network growth mechanism and show that recommendation systems based on clustering and homophily can substantially affect the spread of complex contagions.
This suggests that platform-mediated exposure can play an important role in shaping diffusion outcomes.

However, most of the existing literature focuses on the influence of direct neighbors. The extent to which nodes at greater distances (e.g., friends-of-friends) independently affect adoption remains less well understood, particularly in large-scale online social networks. This motivates the analysis in the present work.

\subsection{Measuring indirect influence in social networks}

The previous subsection reviewed how an individual's immediate neighborhood, i.e., the nodes located at distance $d=1$ from the focal node, affects the node’s behavior. A natural next question is whether nodes located farther away in the network can also shape the focal node’s behavior. This subsection focuses on studies that investigate such indirect influence in social networks.

In a study of public goods games, Fowler and Christakis~\cite{fowler2010cooperative} found that cooperative behavior cascades up to three degrees of separation in the resulting interaction networks, an effect they argued cannot be attributed solely to homophily. This result is closely related to the ``three degrees of influence'' hypothesis, according to which behavioral effects may extend beyond direct social ties to friends-of-friends and even third-degree contacts~\cite{christakis2013social}. Indirect effects have also been found outside human social networks. Guimar{\~a}es et al.~\cite{guimaraes2017indirect} studied ecological interaction networks and showed that non-interacting species can play an important role in trait dynamics, with indirect effects propagating through multiple network pathways and, in some ecological systems,
being as important as direct interactions. More recently, Miranda et al.~\cite{Miranda2024} provided experimental support for indirect influence in the diffusion of innovations. Across 21 experimental sessions with more than 590 participants, they found that adoption was influenced not only by direct neighbors but also by second- and third-degree contacts, with influence decaying with network distance. However, these findings were obtained in a controlled laboratory setting rather than in naturally occurring social media behavior.

Zhang et al.~\cite{zhang2018} examined direct and indirect peer influence in a large-scale mobile carrier network, studying the adoption of caller ringback tones among millions of users. They distinguished effects of direct contacts ($d=1$) from those operating through ``structurally equivalent'' peers, i.e., users who share common ties with the focal individual, and found that both contribute significantly to adoption decisions, although their relative magnitudes vary with neighborhood size. VanderWeele~\cite{vanderWeele2013} developed a statistical framework for studying influence across multiple degrees of separation in social networks, highlighting the methodological challenges of estimating indirect influence while accounting for statistical dependence among connected individuals. While these studies confirm the relevance of indirect peer effects and clarify the methodological challenges of identifying them, they do not directly address social media platforms with algorithmic content curation, where additional exposure pathways may arise.

Xie et al.~\cite{xie2022indirect} analyzed collaboration networks of scientists and demonstrated that indirect influence can, in some settings, be as strong as or even stronger than direct influence. To capture this effect, they introduced the notion of the \emph{induced index}, defined as the maximal number of active neighbors among the active neighbors of a focal node. This quantity reflects the extent to which a node receives reinforcement through second-order connections and provides a natural measure of indirect influence. Motivated by these observations, the authors proposed an induced percolation model that reproduces the empirical patterns observed in the data. The analysis of this model revealed rich phenomenology, including multiple percolation transitions whose nature depends on structural properties of the network, such as reciprocity; for instance, discontinuous (first-order) transitions emerge in networks with non-reciprocal edges.

While this framework provides an elegant theoretical description of indirect influence, its empirical validation has so far been limited to collaboration networks. Moreover, the induced index focuses on second-order effects conditioned on active neighbors and therefore does not capture how active nodes are distributed across the wider local network around the focal node, including second-order activity that is connected to the focal node through inactive immediate neighbors. In the present work, we build on this approach by introducing an extended induced index, denoted $m^*$, which generalizes the original definition by considering all neighbors of the focal node. This modification captures indirect exposure pathways that are invisible to the original index, namely those mediated by inactive immediate neighbors.

Despite this growing body of work, empirical studies of indirect influence on large-scale online social media platforms remain scarce. The present study addresses this gap by examining the spreading patterns of online liking behavior in VKontakte, with a specific focus on the role of friends-of-friends and on the explanatory power of the extended induced index $m^*$.

\section{Methods} \label{Methods}

This section describes the empirical setting (Subsect.~\ref{SSec:Context}) and the dataset (Subsect.~\ref{SSec:Data}), and introduces the induced percolation indices used in the analysis (Subsect.~\ref{SSec:Indices}).

\subsection{Empirical context} \label{SSec:Context}

We gathered data from the Russian social network VKontakte (VK), whose core functionality resembles that of Facebook.
VK is one of the dominant social networking platforms in Russia and Russian-speaking countries; VK reported an average monthly audience of approximately 93 million users in Russia as of Q4 2025\footnote{Reported in VK's financial statements for 2025, \url{https://corp.vkcdn.ru/media/files/ENG_Press_Release_12M_2025.pdf} (accessed 31-05-2026).}.
On VK, users can establish unilateral connections (subscriptions) and dyadic connections (friendships) with other users, and can subscribe to online communities (groups).
Users receive information through a news-feed interface that organizes incoming content as a sequence of items. A user may observe posts or reposts published by their immediate neighbors in the network and may also encounter posts liked or commented on by their friends through the platform's content-delivery mechanisms. Thus, content produced by a second-degree contact may become visible to the focal user when an immediate neighbor interacts with it.

In the present study, we are interested in how liking behavior unfolds on VK. To this end, we first outline the mechanisms through which such behavior may propagate. Formally, consider a post $P$ that has been liked by a set $\mathcal{I}$ of users, and a user $i \notin \mathcal{I}$. For $i$ to like post $P$, two conditions must be met: (i) $i$ must be informed about the post, and (ii) $i$ must decide to perform this action. Prior work (Section~\ref{Sec:Literature}) suggests that the activity of immediate neighbors of $i$ may affect the decision to like. Moreover, VK can inform users about the actions of their friends -- user $i$ can see in their news feed that a friend $j$ has liked or reposted the target post.\footnote{On VKontakte, when a user reposts a post, they implicitly like the original post.}

A key open question is whether more distant connections also affect the probability that user $i$ likes the post. Since the internal ranking rules of VK are not observed, we treat this mechanism as an empirical question: is user $i$ more likely to like post $P$ when more of $i$'s second-degree contacts have liked it, holding direct-neighbor activity fixed? If such an association is observed, it would be consistent with platform-mediated exposure pathways extending beyond direct ties.

An alternative explanation must, however, be acknowledged. Increasing the number of $i$'s second-degree contacts in $\mathcal{I}$ also tends to increase the total number of users who have liked the post, thereby raising its overall popularity. If VK's ranking system is subject to popularity bias, a post's overall popularity may itself increase its visibility, regardless of the network distance between the focal user and the active set. Since we do not observe users' actual feeds or time-varying ranking scores, this mechanism cannot be fully separated from second-order exposure in the present data.

\subsection{Data}\label{SSec:Data}

VK's API allowed us to track likes on a fixed set of posts for users from a predefined sample, as well as friendship ties between users. To obtain a relatively closed social system, we focused on users located in Tomsk Oblast, a region in Siberia, Russia. Between August 15, 2025 and September 15, 2025, we extracted friendship connections among all weekly active VK users with open profiles labeled as belonging to Tomsk Oblast, obtaining a total of $n = 313{,}598$ users. We also collected users' gender.

For each user, we obtained the set of their friends. Since we did not extend the initial sample, friends located outside this set were excluded from the analysis. This yielded $24{,}112{,}008$ outgoing edges ($76.9$ per user). After symmetrization, the sum of degrees in the resulting undirected network rose to $24{,}950{,}496$ (or $79.6$ per user), corresponding to $12{,}475{,}248$ undirected edges. This increase may be due to privacy settings that allow users to hide specific social ties on VK. The resulting undirected social network was disconnected: the giant connected component contained approximately $93.6\%$ of all nodes ($n = 293{,}410$), while the second-largest component contained only 47 nodes. We focused on the giant connected component, which may help reduce the influence of inactive or anomalous accounts located at the network periphery~\cite{gonzalez2021bots}.

\begin{figure}[tb]
	\centering
	\includegraphics[width=0.5\linewidth]{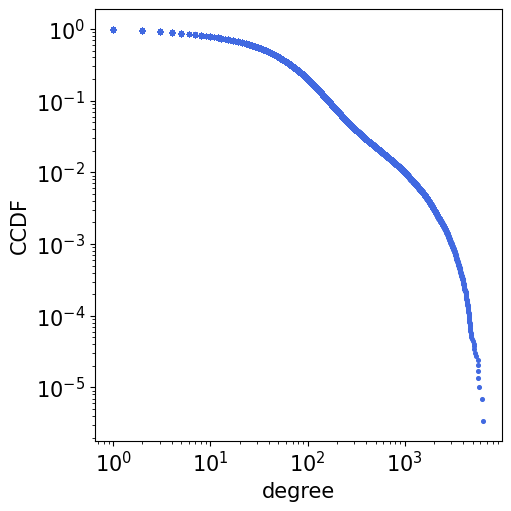}
	\caption{CCDF for the degree distribution of the social network.
    }\label{Fig - CCDF}
\end{figure}

The resulting giant component is highly clustered. Its transitivity coefficient is $C = 0.138$, compared with $C_0 \approx 2.9 \times 10^{-4}$ for an Erd{\"o}s--R{\'e}nyi random graph with the same number of nodes and edges. The average path length is $\langle d \rangle \approx 3.42$, indicating a small-world structure, i.e., short typical distances combined with high clustering~\cite{watts1998collective}. The average degree is $\langle k \rangle = 85.03$ (standard deviation $218.95$), and the maximal degree is $k_{\max} = 6393$. Figure~\ref{Fig - CCDF} shows the complementary cumulative distribution function (CCDF) of the degree distribution, which departs from a power-law pattern and exhibits a steep drop starting at approximately $k \approx 10^3$.

Using these data, we traced how liking behavior spreads among VK friends. To this end, we selected a list of $N = 308$ posts published by Tomsk local news outlets, which primarily cover local issues. The posts were chosen to have either at least 20 likes or at least 10 reposts. Due to API limitations, we focused our data-collection capacity on relatively popular posts; this choice may introduce a selection bias, which we discuss later. For each post, the VK API allows to get a chronological list of likes. In total, we identified $L = 36{,}286$ likes. Of these, $19{,}599$ likes were made by users in the giant connected component, meaning that approximately $46\%$ of likes came from users outside the analyzed component.

These data allow us to investigate how the probability of liking a post changes with the number of active nodes, corresponding to the users who have already liked the post, at various distances from the focal user, including immediate friends ($d = 1$) and friends-of-friends ($d = 2$).

\subsection{Second-order exposure indices}\label{SSec:Indices}

Our main objective is to understand the role of nodes at distance $d=1$ and, especially, at distance $d=2$ in shaping liking behavior. To this end, we consider a focal node $i$ that has not yet liked a given post and examine how the probability of liking depends on the number of immediate friends who have already liked the post, i.e., the active neighbors of the focal node, denoted by $k_1$, and on the number of active nodes at distance $d=2$, denoted by $k_2$. At the time of measurement, the focal node is not active and therefore is not counted among anyone's active neighbors.  Following~\cite{xie2022indirect}, we first consider the \emph{induced index} $m$, defined as the largest number of active neighbors among the \emph{active} immediate neighbors of the focal node. We then introduce the \emph{extended induced index} $m^*$, defined analogously but with the maximum taken over \emph{all} immediate neighbors of the focal node, regardless of whether they are themselves active.

\begin{figure}[t!]
	\centering
	\includegraphics[width=0.75\linewidth]{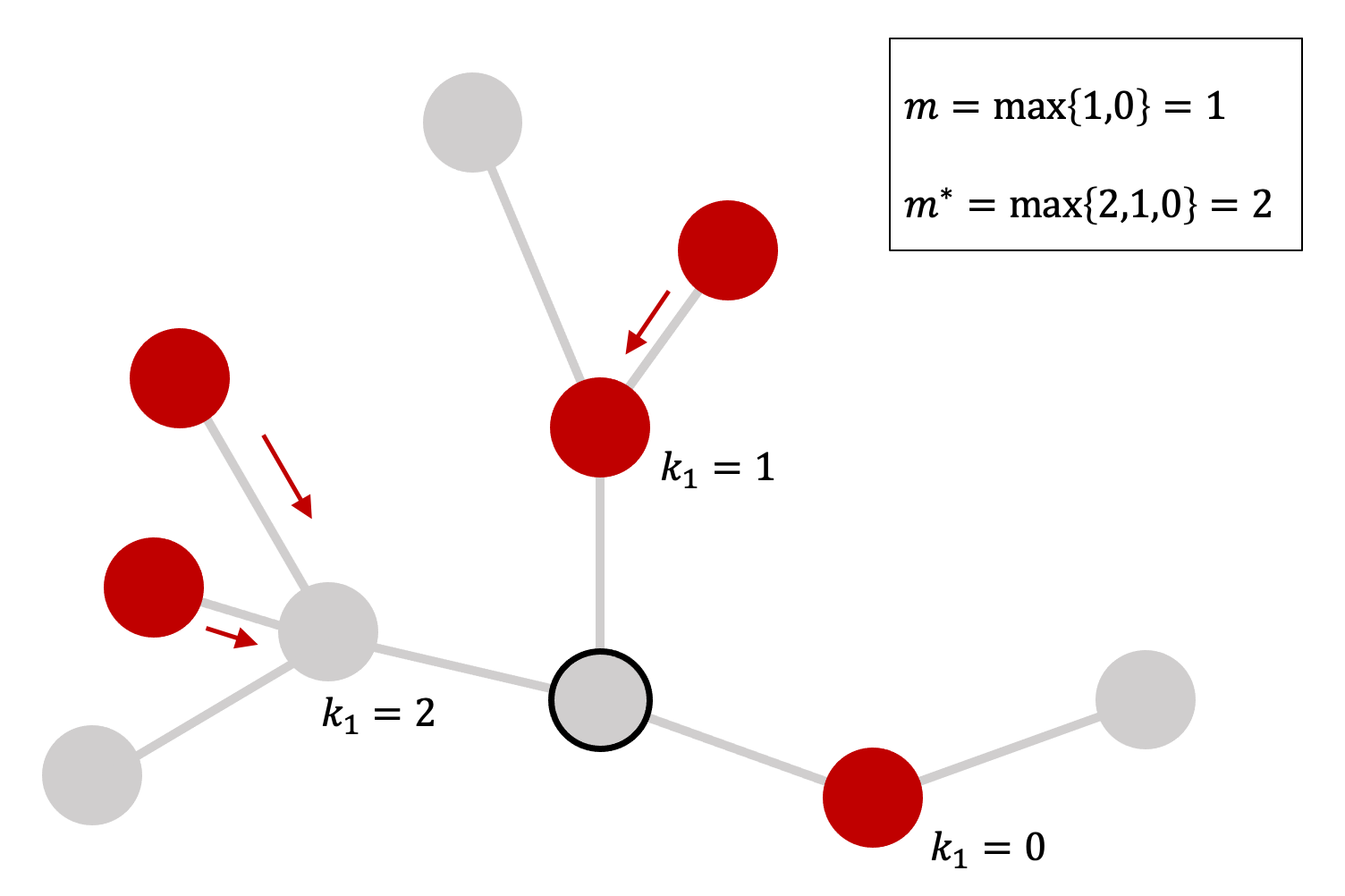}
	\caption{The induced indices $m$ and $m^*$ operationalize $d=2$ influence by recording the maximum number of active
friends (highlighted in maroon) over the immediate neighbors of a focal node (indicated by a black outline).
In this example, the inactive focal node has three neighbors, two of which are active.
For $m$ (induced index), the maximum is taken over the active neighbors of the focal node; in this example, these neighbors have $0$ and $1$ active neighbors, respectively, so $m=1$. For $m^*$ (extended induced index),
the maximum is taken over all neighbors of the focal node; hence, $m^*=2$.
}\label{Fig: Induced indices}
\end{figure}

In our analysis, we treat each post as an independent information cascade, which induces its own distribution of $d=1$ and $d=2$ active neighborhoods. We focus on observations in which the structural diversity index $s$, defined as the number of connected components among active neighbors~\cite{ugander2012structural}, does not exceed four. The excluded observations constitute less than $0.001\%$ of the data and represent extreme outliers in the structural diversity distribution. The resulting dataset contains $N = 90{,}343{,}268$ observations (one observation per user–post pair), allowing for high-precision estimation. In the analysis below, we focus on this subset of the data.

\section{Results} \label{Sec:Results}

We present the main results of this paper in three steps. We first examine the association between liking behavior and activity among direct friends ($d=1$), then turn to second-order activity ($d=2$), and finally assess the joint explanatory power of the proposed indices using regression models.

For each metric of interest (e.g., $k_1$, $k_2$, $m^*$), we estimate the conditional probability of liking as the fraction of
user-post pairs in which the user liked the post, among all pairs sharing the same value of the conditioning variable(s). Because the baseline liking probability is very low, we report relative probabilities, normalized by the probability of liking for users with no active friends ($k_1 = 0$). Unless otherwise stated, we apply the chronological filtering: for users who liked the post, the active neighborhood is measured at the instant immediately before the focal user's like, so that only preceding likes are counted. For users who did not like the post, the active neighborhood is computed using all likes recorded in the observation period.


\subsection{$d=1$ influence}

\begin{figure}[t!]
	\centering
	\includegraphics[width=\linewidth]{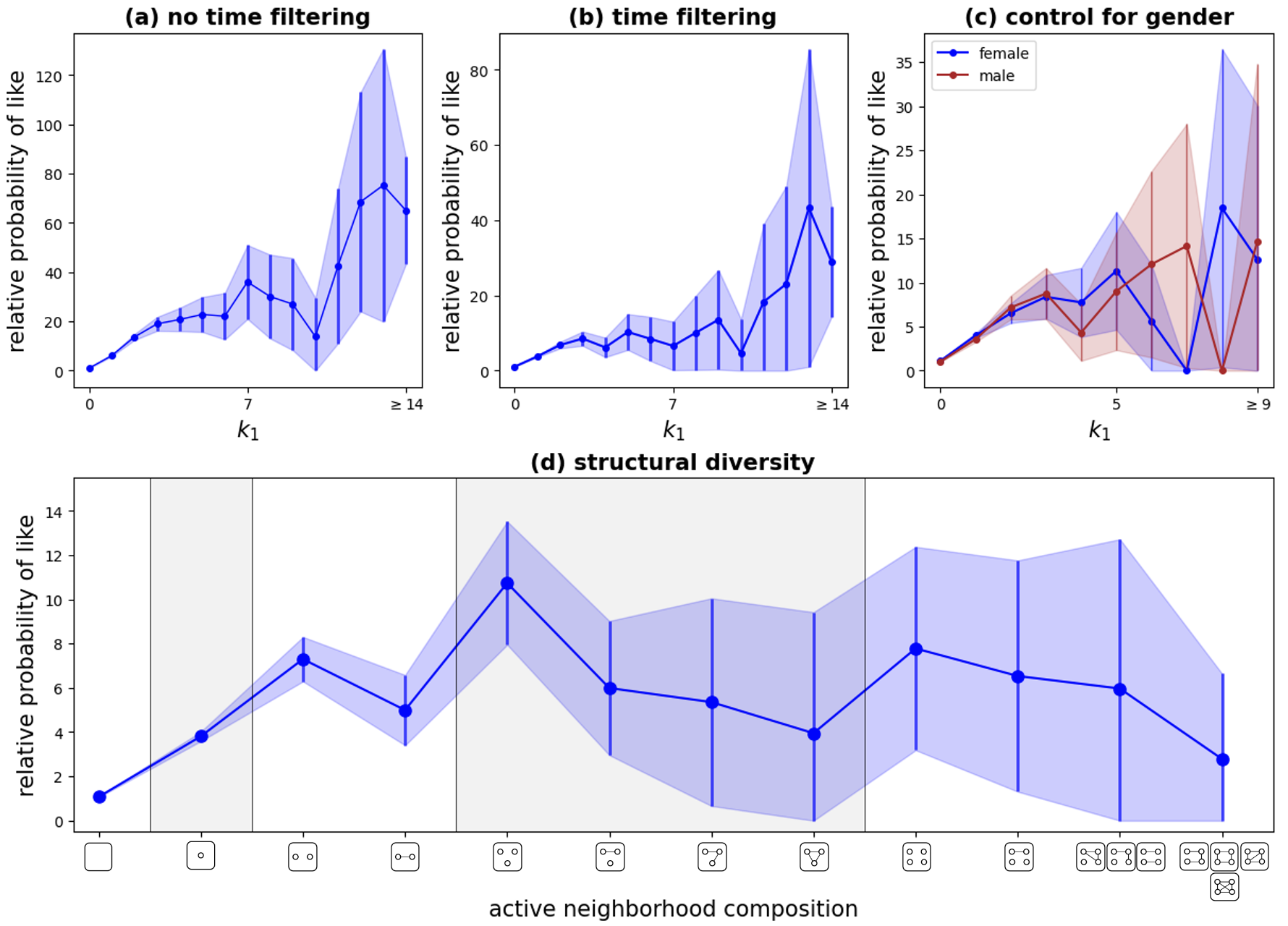}
	\caption{Relative probability of liking as a function of the user's immediate neighborhood. Panel (a) shows the effect of the number of active friends; panel (b) shows the same quantity after applying the chronological filter. Panel (c) separates the results by gender. Panel (d) shows the effect of structural diversity, measured by the number of connected components among active friends. All probabilities are normalized by the probability of liking for users with no active friends ($k_1=0$), so the corresponding value equals one.}\label{Fig:Peer_effect_(d=1)}
\end{figure}

We start by examining the role of nodes at distance $d=1$. Figure~\ref{Fig:Peer_effect_(d=1)} shows how the relative probability of liking depends on the number of active friends and on the structural diversity index. Relative probabilities are normalized by the probability of liking for users with no active friends ($k_1=0$), so the corresponding value equals one. Panels (a) and (b) show that the probability of liking increases with the number of active users in the immediate neighborhood, consistent with previous empirical work on social contagion. Having one active friend increases the probability of liking by approximately a factor of four relative to the case with no active friends. For panel (a), however, chronological filtering has not been applied. After imposing chronological ordering and counting only likes that occurred before the focal user's like, the overall level of the estimated probabilities decreases, but the monotonically increasing and concave trend for small $k_1$ remains. Panels (b)--(d) in Figure~\ref{Fig:Peer_effect_(d=1)} and all subsequent analyses apply this chronological filter.

Panel (c) shows no clear gender difference in the association between active friends and liking probability: males and females exhibit very similar response curves. However, gender-related effects may become visible after accounting for the gender composition of the active neighborhood, as demonstrated in a different behavioral domain by Aral and Nicolaides~\cite{aral2017exercise}. Panel (d) provides evidence consistent with structural diversity: for a fixed number of active friends, the probability of liking is higher when these friends are distributed across a larger number of connected components.

\begin{figure}[t!]
	\centering
	\includegraphics[width=0.9\linewidth]{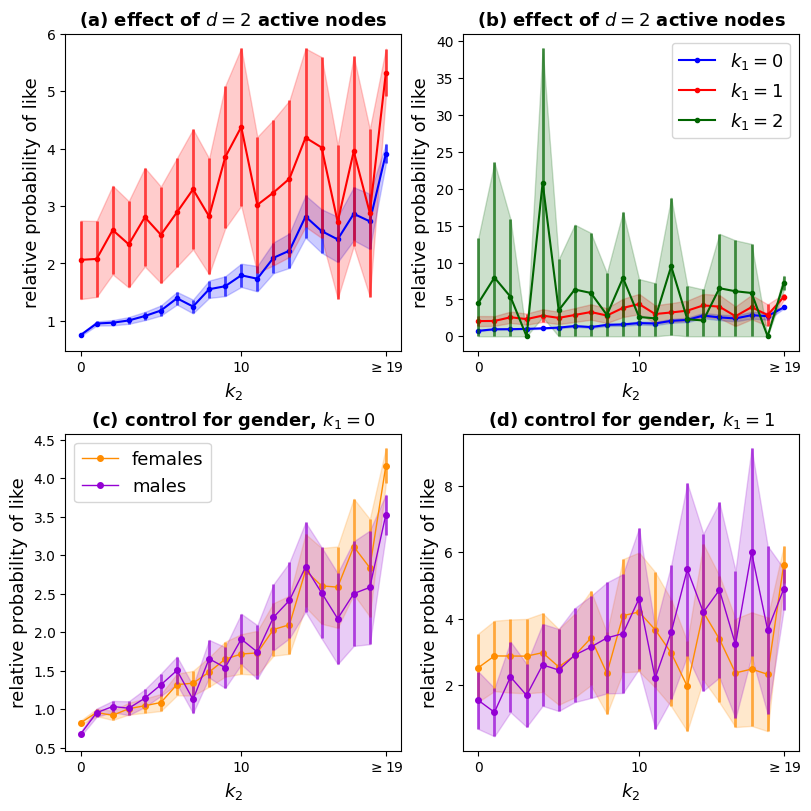}
	\caption{Relative probabilities of liking as functions of the number of active $ d=2 $ nodes. In panels (a) and (b) we control for the number of active friends $ k_1 $, panels (c) and (d) control for both $ k_1 $ and the gender of the focal node. The normalization value is the same as in Figure~\ref{Fig:Peer_effect_(d=1)}.}\label{Fig:Peer_effect_(d=2)_part_1}
\end{figure}

\subsection{$d=2$ influence} \label{Sec:d=2 influence}

We next examine activity at distance $d=2$. Panel (a) of Figure~\ref{Fig:Peer_effect_(d=2)_part_1} shows that the probability of liking increases with the number of active friends-of-friends. We plot this relationship separately for users with no active direct friends ($k_1=0$, blue curve) and for users with one active direct friend ($k_1=1$, red curve). In both cases, the probability of liking increases with $k_2$, suggesting that second-order activity remains informative even after stratifying users by the number of active direct friends. For users with more than one active friend, the estimates become noisier (panel (b)), and we refrain from drawing strong conclusions for this regime. Panel (c) shows no clear gender differences, whereas panel (d) suggests that, among users with one active friend, females have a higher probability of liking than males for $k_2 \leq 5$.
The relatively narrow confidence intervals observed for $k_2 \geq 19$ in panels (a) and (b) of Figure~\ref{Fig:Peer_effect_(d=2)_part_1} are explained by the large number of observations in this range. Because the network has short typical distances (recall that $\langle d \rangle \approx 3.42$), many users have a non-negligible number of active nodes at distance two.

\begin{figure}[t!]
	\centering
	\includegraphics[width=\linewidth]{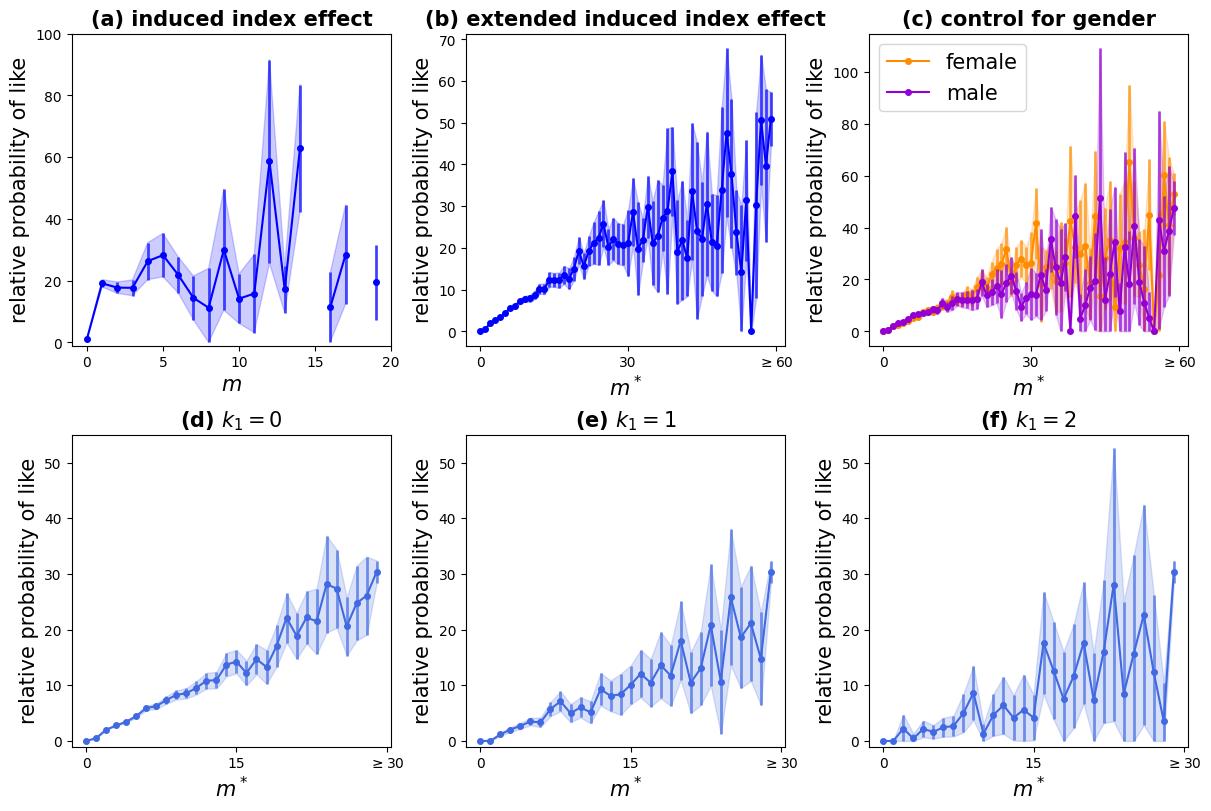}
	\caption{Relative probability of liking as a function of the induced indices. Panel (a) shows the induced index $m$, while panel (b) shows the extended induced index $m^*$. Panel (c) separates the results by gender. Panels (d)--(f) show the relationship for users with no active friends, one active friend, and two active friends, respectively. The normalization value is the same as in Figure~\ref{Fig:Peer_effect_(d=1)}.}\label{Fig:Peer_effect_(d=2)_part_2}
\end{figure}

The induced index $m$ is positively associated with the probability of liking (Figure~\ref{Fig:Peer_effect_(d=2)_part_2}, panel (a)). The increase from $m=0$ to $m=1$ is accompanied by a sharp rise in the relative probability of liking, by approximately a factor of 19. This appears to be the largest one-step increase observed in the low-value range of our peer-effect metrics (cf. Figures~\ref{Fig:Peer_effect_(d=1)} and~\ref{Fig:Peer_effect_(d=2)_part_1}). After this initial jump, the curve levels off, with another increase around $m=3$, where the relative probability approaches approximately 30. For larger values of $m$, the estimates become too noisy to draw firm conclusions.

The extended induced index $m^*$ exhibits a much clearer pattern. As shown in Figure~\ref{Fig:Peer_effect_(d=2)_part_2}, panel (b), the probability of liking increases near-linearly with $m^*$. We also observe possible interaction patterns involving $m^*$, gender, and the number of active friends. Panel (c) suggests that, around $m^* \approx 27$, females have a higher probability of liking than males. Panels (d)--(f) show that, after stratifying by $m^*$, larger values of $k_1$ are associated with a lower probability of liking. The regression analysis in Section~\ref{Sec:RegAnalysis} supports this surprising conditional negative association and also provides further evidence for the positive role of structural diversity observed in Figure~\ref{Fig:Peer_effect_(d=1)}.

These patterns suggest that the raw number of active friends $k_1$ is not by itself sufficient to characterize network effects, because structural diversity and second-order activity carry additional explanatory information. In the regression analysis reported below, the conditional association between $k_1$ and liking probability changes sign once these variables are included.

Motivated by the three-degrees-of-influence hypothesis~\cite{fowler2010cooperative}, we also examined activity at distances $d \geq 3$. We did not detect a clear additional association between such distant active nodes and the probability of liking once activity at distances $d=1$ and $d=2$ was considered. This finding does not refute the three-degrees hypothesis, since our analysis does not reconstruct full temporal propagation paths. Rather, it indicates that, in the present data, the strongest detectable nonlocal signal is concentrated at distance $d=2$. Importantly, second-order activity remains informative even in the absence of active direct friends: when $k_1=0$, both $k_2$ and $m^*$ are positively associated with liking probability (Figure~\ref{Fig:Peer_effect_(d=2)_part_1}, panel (a), and Figure~\ref{Fig:Peer_effect_(d=2)_part_2}, panel (d)).

\subsection{Regression analysis} \label{Sec:RegAnalysis}

To confirm our findings and explore the joint effects of several predictors, we estimate a sequence of logistic regression models:
\begin{equation} \label{eq:logit}
    P(\text{like} \mid x) =
    \frac{1}{1 + \exp[-(\beta_0 + \beta^\top x)]}
    = \pi(\beta_0 + \beta^\top x),
\end{equation}
where $x$ is the vector of explanatory variables, $\beta$ is the corresponding coefficient vector, $\beta_0$ is the intercept, and $\pi(z) = (1+\exp(-z))^{-1}$ denotes the logistic link function.

\paragraph{Model sequence and covariates}

We start with the simplest model, which includes only the number $k_1$ of active friends ($\mathcal{M}_1$) and therefore captures the basic threshold-like effect of direct exposure:
\[
P(\text{like} \mid x) = \pi(\beta_0 + \beta_1 k_1).
\]
Each subsequent model either adds one explanatory variable or replaces a variable with an alternative operationalization, allowing us to compare their contributions to model fit (see Tables~\ref{tab:RegressionFit} and~\ref{tab:RegressionFit_add}). In particular, model $\mathcal{M}_2$ adds the density $\rho_1$ of ties among active friends (i.e., the proportion of realized edges in the subgraph induced by active neighbors). Model $\mathcal{M}_3$ adds the structural diversity index $s$. Starting from model $\mathcal{M}_4$, we incorporate predictors based on activity at distance $d=2$. Table~\ref{tab:RegressionFit_add} continues the model sequence after $\mathcal{M}_7$ by considering additional covariates.

We consider several formalizations of this second-order activity: the number $k_2$ of active nodes at distance $d=2$, the induced indices $m$ and $m^*$, and two discretized versions of the induced index $m$. This choice is motivated by the threshold-like pattern discussed in Section~\ref{Sec:d=2 influence}: the increases from $m=0$ to $m=1$ and from $m=3$ to $m=4$ are associated with particularly sharp increases in the probability of liking, among the largest one-step changes observed in our data.

This suggests that the effect of $m$ may be better captured by threshold indicators than by a single linear term. We therefore define two indicator variables: $\mathbbm{1}_{1\leq m\leq 3}$, which equals one if $1\leq m\leq 3$ and zero otherwise, and $\mathbbm{1}_{m>3}$, which equals one if $m>3$ and zero otherwise. Using these two indicators instead of the raw value of $m$ allows us to capture the stepwise shape of the empirical $m$-curve.

Since the extended induced index specification $\mathcal{M}_6$ provides the best fit in Table~\ref{tab:RegressionFit}, we repeat it as a reference baseline before adding gender and degree and before comparing it with the stepwise-indicator specification.

\paragraph{Multicollinearity diagnostics}
Before estimating the regression models, we examined dependencies among the explanatory variables using the variance inflation factor (VIF). We found that the number of active friends $k_1$ (VIF = $73.44$) and the structural diversity index $s$ (VIF = $67.3$) are strongly correlated with the remaining variables, and in particular with each other\footnote{Notice that such a correlation can be expected, since $s\leq k_1$ by construction.}, whereas the other predictors have VIF values in an acceptable range (Table~\ref{tab:VIF}). If either $k_1$ or $s$ is removed, the VIF values of the remaining variables fall below $2.5$, as shown in the second and third columns of the table. We retain both $k_1$ and $s$ in the main models because they capture different aspects of exposure to active direct neighbors. Since these two variables are highly correlated, however, their individual coefficient estimates should be interpreted with caution. As detailed below, nearly all estimated effects appeared to be statistically significant at $P<0.001$, which outcome is expected given the large sample size. We therefore focus on the direction of effects, their stability across model specifications, and overall model fit rather than on statistical significance alone.

\begin{table}[t]
    \renewcommand{\arraystretch}{2}
    \centering
    \caption{Variance inflation factor (VIF) analysis for the main explanatory variables}
    \label{tab:VIF}
    \begin{tabular}{>{\raggedright\arraybackslash}p{0.35\linewidth}>{\raggedright\arraybackslash}p{0.15\linewidth}>{\raggedright\arraybackslash}p{0.15\linewidth}>{\raggedright\arraybackslash}p{0.15\linewidth}}
        \hline
        &All variables& excluding $k_1$ & excluding $s$ \\
        \hline
        intercept & 2.49 & 2.49 & 2.49  \\
        number of active friends $k_1$ & \textbf{73.44} & \textbf{1.35} &   \\
        density $\rho_1$ of active $d=1$ network & 3.09 & 1.12 & 1.04  \\
        structural diversity index $s$ & \textbf{67.3} &  & \textbf{1.22}  \\
        extended induced index $ m^* $ & 1.76 & 1.82 & 1.75  \\
        gender (1-male,0-female) & 1 & 1 & 1  \\
        degree & 1.17 & 1.17 & 1.17  \\
        \hline
    \end{tabular}
\end{table}

\paragraph{Model comparison}
To compare model specifications, we use the Bayesian Information Criterion (BIC) and McFadden's pseudo-$R^2$, two standard likelihood-based measures for categorical-response models~\cite{long2021regression}. BIC balances goodness of fit against model complexity and imposes a sample-size-dependent penalty for additional parameters. Pseudo-$R^2$ compares the log-likelihood of the fitted model with that of the intercept-only model.
Although McFadden's pseudo-$R^2$ values remain modest in absolute terms, this is expected for rare-event behavioral outcomes at the user-post level. We therefore interpret pseudo-$R^2$ primarily as a comparative measure across model specifications rather than as an absolute measure of explained variance.
Our regression analysis (Tables~\ref{tab:RegressionFit} and~\ref{tab:RegressionFit_add}) reveals several distinct coefficient patterns.
\begin{table}[t]
    \renewcommand{\arraystretch}{3}
    \tiny
    \centering
    \caption{Regression fit (part 1); coefficient estimates are reported along with standard errors (in parentheses).}
    \label{tab:RegressionFit}
    \begin{tabular}{>{\raggedright\arraybackslash}m{0.09\linewidth}>{\centering\arraybackslash}m{0.07\linewidth}>{\centering\arraybackslash}m{0.08\linewidth}>{\centering\arraybackslash}m{0.08\linewidth}>{\centering\arraybackslash}m{0.08\linewidth}>{\centering\arraybackslash}m{0.08\linewidth}>{\centering\arraybackslash}m{0.08\linewidth}>{\centering\arraybackslash}m{0.1\linewidth}}
        \hline

        Variable
        & \makecell[c]{$\mathcal{M}_1$\\[1pt]{baseline}}
        & \makecell[c]{$\mathcal{M}_2$\\[1pt]{$\mathcal{M}_1+\rho_1$}}
        & \makecell[c]{$\mathcal{M}_3$\\[1pt]{$\mathcal{M}_2+s$}}
        & \makecell[c]{$\mathcal{M}_4$\\[1pt]{$\mathcal{M}_3+k_2$}}
        & \makecell[c]{$\mathcal{M}_5$\\[1pt]{$\mathcal{M}_3+m$}}
        & \makecell[c]{$\mathcal{M}_6$\\{$\mathcal{M}_3+m^*$}}
        & \makecell[c]{$\mathcal{M}_7$\\[1pt]{$\mathcal{M}_3$+indicators}} \\

        \hline

        Intercept & \makecell[c]{-8.4\\ (0.0072)} & \makecell[c]{-8.4\\ (0.0072)} & \makecell[c]{-8.5\\ (0.0073)} & \makecell[c]{-8.5\\ (0.0074)} & \makecell[c]{-8.5\\ (0.0073)} & \makecell[c]{-8.7\\ (0.0077)} & \makecell[c]{-8.5\\ (0.0075)} \\

        Active friends $k_1$ & \makecell[c]{0.14\\ (0.0039)} & \makecell[c]{0.13\\ (0.0040)} & \makecell[c]{-0.38,\\ \textbf{P=0.03}\\ (0.18)} & \makecell[c]{-1.0\\ (0.18)} & \makecell[c]{-1.2\\ (0.19)} & \makecell[c]{-1.2\\ (0.18)} & \makecell[c]{-0.21,\\ \textbf{P=0.137}\\ (0.14)} \\

        Density of active $d = 1$ network $\rho_1$ & & \makecell[c]{1.3\\ (0.14)} & \makecell[c]{0.99\\ (0.27)} & \makecell[c]{1.2\\ (0.27)} & \makecell[c]{1.2\\ (0.28)} & \makecell[c]{0.44,\\ \textbf{P=0.114}\\ (0.28)} & \makecell[c]{-1.7\\ (0.25)} \\

        Structural diversity $s$ & & & \makecell[c]{1.3\\ (0.18)} & \makecell[c]{1.5\\ (0.18)} & \makecell[c]{2.0\\ (0.19)} & \makecell[c]{1.1\\ (0.18)} & \makecell[c]{-2.1\\ (0.17)} \\

        Active friends-of-friends $k_2$ & & & & \makecell[c]{0.0086\\ (0.00023)} & & & \\

        Induced index $m$ & & & & & \makecell[c]{0.087\\ (0.0021)} & & \\

        Extended induced index $m^*$ & & & & & & \makecell[c]{0.083\\ (0.00050)} & \\

        Indicator $\indi{1 \leq m \leq 3}$ & & & & & & & \makecell[c]{5.5\\ (0.088)} \\

        Indicator $\indi{m>3}$ & & & & & & & \makecell[c]{6.3\\ (0.098)} \\

        \hline
        McFadden’s pseudo-$R^2$ & 0.0013 & 0.0015 & 0.0048 & 0.0075 & 0.0074 & \textbf{0.0384} & 0.0258 \\
        BIC & 369,418.58 & 369,381.59 & 367,070.85 & 366,115.88 & 366,155.65 & \textbf{354,723.10} & 359,371.78 \\


        \hline
    \end{tabular}
    \begin{tablenotes}
        \item[] Unless an exact $P$-value is indicated, all estimates are statistically significant at $P<0.001$.
        \item[] The best goodness-of-fit is highlighted in bold.
    \end{tablenotes}
\end{table}

\begin{table}[h]
\renewcommand{\arraystretch}{3}
\tiny
\centering
\caption{Regression fit (part 2); coefficient estimates are reported along with standard errors (in parentheses). The fit of model $\mathcal{M}_6$ is replicated from Table~\ref{tab:RegressionFit} as a reference.}
\label{tab:RegressionFit_add}
\begin{tabular}{>{\raggedright\arraybackslash}m{0.1\linewidth}>{\centering\arraybackslash}m{0.14\linewidth}>{\centering\arraybackslash}m{0.14\linewidth}>{\centering\arraybackslash}m{0.14\linewidth}>{\centering\arraybackslash}m{0.14\linewidth}}
\hline



    \textbf{Variable }
        & \makecell[c]{$\mathcal{M}_6$\\[1pt]{baseline}}
        & \makecell[c]{$\mathcal{M}_8$\\[1pt]{$\mathcal{M}_6$+gender}}
        & \makecell[c]{$\mathcal{M}_9$\\[1pt]{$\mathcal{M}_8$+degree}}
        & \makecell[c]{$\mathcal{M}_{10}$\\[1pt]{$\mathcal{M}_9$ replacing $m^*$}\\{with indicators}}\\

\hline

    Intercept & \makecell[c]{-8.7\\ (0.0077)} & \makecell[c]{-8.6\\ (0.0099)} & \makecell[c]{-8.7\\ (0.010)} & \makecell[c]{-8.5\\ (0.0100)} \\

    Active friends $k_1$ & \makecell[c]{-1.2\\ (0.18)} & \makecell[c]{-1.2\\ (0.18)} & \makecell[c]{-1.2\\ (0.18)} & \makecell[c]{-0.25\\ \textbf{P=0.081}\\ (0.14)} \\

    Density of active $d=1$ network $\rho_1$ & \makecell[c]{0.44\\ \textbf{P=0.114}\\ (0.28)} & \makecell[c]{0.44\\ \textbf{P=0.116}\\ (0.28)} & \makecell[c]{0.38\\ \textbf{P=0.172}\\ (0.28)} & \makecell[c]{-1.7\\ (0.25)} \\

    Structural diversity $s$ & \makecell[c]{1.1\\ (0.18)} & \makecell[c]{1.1\\ (0.18)} & \makecell[c]{1.1\\ (0.18)} & \makecell[c]{-2.1\\ (0.17)} \\

    Extended induced index $m^*$ & \makecell[c]{0.083\\ (0.00050)} & \makecell[c]{0.083\\ (0.00050)} & \makecell[c]{0.083\\ (0.00050)} & \\

    Indicator $\indi{1 \leq m \leq 3}$ & & & & \makecell[c]{5.5\\ (0.088)} \\

    Indicator $\indi{3 < m}$ & & & & \makecell[c]{6.3\\ (0.097)} \\

    Male gender & & \makecell[c]{-0.051\\ (0.015)} & \makecell[c]{-0.050\\ (0.015)} & \makecell[c]{-0.079\\ (0.015)} \\

    Degree & & & \makecell[c]{0.00032\\ (0.000023)} & \makecell[c]{0.00011\\ (0.000025)} \\

    \hline

    McFadden’s pseudo-$R^2$ & 0.0384 & 0.0384 & \textbf{0.0388} & 0.0259 \\

    BIC & \makecell[c]{354,723.10} & \makecell[c]{354,729.20} & \makecell[c]{\textbf{354,593.39}} & \makecell[c]{359,360.60} \\


    \hline
\end{tabular}
\begin{tablenotes}
    \item[] Unless an exact $P$-value is indicated, all estimates are statistically significant at $P<0.001$.
    \item[] The best goodness-of-fit is highlighted in bold.
\end{tablenotes}
\end{table}

All $d=2$ metrics are positively associated with the liking probability. Among the continuous second-order measures, the extended induced index $m^*$ provides the strongest improvement in model fit, yielding the lowest BIC and the highest McFadden pseudo-$R^2$ in Table~\ref{tab:RegressionFit}. In these specifications, the number of active friends $k_1$ is \emph{negatively} associated with liking probability once structural diversity $s$ and the density $\rho_1$ of ties among active friends are included. At the same time, including the extended induced index $m^*$ makes the coefficient of $\rho_1$ statistically insignificant, whereas structural diversity consistently improves model fit and remains positively associated with liking probability.

While the stepwise indicators based on $m$ also substantially improve fit relative to models using only $k_1$, $\rho_1$, and $s$, they do not outperform the specification based on $m^*$ in the main regression table (Table~\ref{tab:RegressionFit}).
When the stepwise indicators are included (model $\mathcal{M}_7$ in Table~\ref{tab:RegressionFit} and model $\mathcal{M}_{10}$ in Table~\ref{tab:RegressionFit_add}), the statistical significance of $k_1$ is substantially reduced. In these specifications, $\rho_1$ and $s$ become negatively associated with liking probability, a pattern that is not expected under the usual structural-diversity interpretation.

Adding gender to the regression model (model $\mathcal{M}_8$, Table~\ref{tab:RegressionFit_add}) has little effect on goodness of fit. Nevertheless, male gender is consistently associated with a lower probability of liking, in line with the descriptive patterns in Figure~\ref{Fig:Peer_effect_(d=1)}, panel (c), and Figure~\ref{Fig:Peer_effect_(d=2)_part_2}, panel (c). Nodal degree is positively
associated with liking probability and only slightly improves model fit (model $\mathcal{M}_9$). Thus, high-degree nodes are, on average, more likely to like a post, possibly because they are more active on the platform overall.

\subsection{Robustness checks} \label{Sec:Robustness}

We also performed robustness checks (Tables~\ref{tab:RegressionFit_checks} and~\ref{tab:RegressionFit_checks_(part_2)}).
First, we estimated the models separately for low-degree ($k < k_{33} = 19$) and high-degree nodes ($k > k_{66} = 62$), where $k_{33}$ and $k_{66}$ denote the 33rd and 66th percentiles of the degree distribution, respectively. This split is motivated by the possibility that high-degree nodes may include commercial or organizational accounts whose activity is less socially driven, whereas low-degree nodes may include inactive users or anomalous accounts.

For low-degree nodes, the $d=1$ variables become statistically insignificant, plausibly because active direct neighbors are rare in this subsample. By contrast, the effects of the $d=2$ indices, as well as gender and degree, persist.
For both low-degree and high-degree nodes, the extended induced index $m^*$ is slightly more informative than the stepwise induced-index indicators.
For high-degree nodes, the gender association reverses: male users are associated with a higher probability of liking in model $\mathcal{M}_9$.

Because homophily is a common alternative explanation in network analysis~\cite{tang2025empirical} for observed peer effects, we also examined whether the results depend on local network closure. To this end, we computed the local clustering coefficient $C$ and estimated the models separately for nodes with low-clustering neighborhoods ($C<C_{33}=0.077$) and high-clustering neighborhoods ($C>C_{66}=0.146$), where $C_{33}$ and $C_{66}$ denote the 33rd and 66th percentiles of the clustering-coefficient distribution (Table~\ref{tab:RegressionFit_checks_(part_2)}).

For low-clustering nodes, the results are broadly similar to those obtained for the full sample, both for model $\mathcal{M}_9$ and for model $\mathcal{M}_{10}$. For high-clustering nodes, however, the gender coefficient changes sign: the point estimate for male gender becomes positive, though it is only marginally significant in $\mathcal{M}_9$ ($P=0.040$) and not significant in $\mathcal{M}_{10}$ ($P=0.356$).
Thus, in highly clustered local neighborhoods, male users appear more likely to like a post, whereas in the full sample and in other subsamples the association is negative.





\section{Discussion and concluding remarks}\label{Sec:discuss}

Using high-resolution data from the social platform VKontakte, we examined how liking behavior spreads across the social network. Our main objective was to assess the role of nodes at distance $d=2$ in this process. Prior work by Xie et al.~\cite{xie2022indirect} introduced the framework of induced percolation, which emphasizes the role of active second-order neighbors in behavioral spreading. The term ``active'' is borrowed from the literature on social contagion~\cite{ugander2012structural} and denotes a node that has adopted the target behavior and may contribute to its further diffusion. Xie and co-authors argued that some behaviors may propagate through active second-order neighbors, with immediate neighbors acting as local conduits of reinforcement. This mechanism is captured by the induced index $m$, defined as the maximal number of active neighbors among the active direct ($d=1$) neighbors of a focal node.
Building on this idea, we tested whether induced-percolation mechanisms help explain liking behavior in a large-scale online social media setting and introduced the extended induced index $m^*$, which takes the same maximum over all immediate neighbors of the focal node.

\subsection{Main findings}

Our data show a positive association between activity at distance $d=2$ and the probability of liking. We tested three $d=2$ measures: the raw count $k_2$, the induced index $m$, and the extended induced index $m^*$. The strongest associations are observed for the two induced indices, although their empirical shapes differ. The induced index $m$ displays a clearly stepwise pattern, with sharp increases around $m=1$ and $m=3$. By contrast, the relationship between liking probability and $m^*$ is nearly linear.

We obtain more mixed evidence regarding the role of activity at distance $d=1$. Descriptively, the probability of liking increases with the number of active friends. In the regression models, however, the conditional coefficient of $k_1$ becomes negative once structural diversity and $d=2$ activity are included. This does not necessarily imply that active direct friends suppress liking; rather, it indicates that the raw number of active friends is difficult to interpret once other aspects of the active neighborhood are controlled for. In particular, $k_1$ is strongly correlated with the structural diversity index $s$, which measures the number of connected components among active friends.

The role of the structural diversity index $s$ remains somewhat ambiguous. In the specifications based on $m^*$, $s$ is positively associated with liking probability, which is consistent with the structural diversity hypothesis. However, when the induced index is represented by stepwise indicators, the coefficient of $s$ becomes negative. We therefore interpret the positive association between the number of connected components among active friends and the probability of liking as suggestive rather than definitive, especially given the dependence among $k_1$, $s$, and the induced-index variables. Clarifying the relationship between structural diversity and induced-index-based measures of indirect exposure is an important topic for future research.

Apart from peer-effect indices, we also examined node-level characteristics, namely gender and nodal degree. Degree is positively associated with liking probability, although the magnitude of this association is small. This suggests that users with many friends are, on average, more likely to like posts, possibly because they are more active on the platform overall. Gender is also associated with liking behavior: in the full-sample models, male users are less likely to like a post than female users, all else being equal. However, the robustness checks indicate that this association is not uniform across all network contexts, since it weakens or changes sign in some high-degree and high-clustering subsamples.

Overall, the results indicate that liking probability is jointly associated with activity at distances $d=1$ and $d=2$. At distance $d=1$, the raw number of active friends is positively associated with liking descriptively, but its conditional regression coefficient becomes negative once structural diversity and second-order activity are included. The structural diversity index $s$, measuring the number of connected components among active friends, remains positively associated with liking in the specifications based on $m^*$. At distance $d=2$, the extended induced index $m^*$ has the strongest and most stable association with liking probability. Taken together, these findings provide empirical support for induced-index-based descriptions of indirect influence~\cite{xie2022indirect} and, within the main model specifications, support for the structural diversity hypothesis~\cite{ugander2012structural}.

\subsection{Platform-mediated exposure and local-network structure}

At this point, it is useful to recall that liking a post requires two conditions (see Section~\ref{SSec:Context}): the user must become aware of the post, and the user must decide to like it. The observed association with $d=2$ activity may therefore reflect not only social reinforcement in the usual sense, but also platform-mediated visibility. Under this interpretation, active friends-of-friends do not necessarily influence the focal user directly; rather, their activity may increase the probability that the post is surfaced to the user through VKontakte's ranking or recommendation mechanisms.

This interpretation is consistent with work on social-network recommendation systems, where triadic closure, common-neighbor structure, and homophily can be used as recommendation signals~\cite{carullo2015triadic}. It is also consistent with recent modeling work showing that link-recommendation mechanisms can reshape network structure and affect complex contagion dynamics~\cite{komander2025modeling}. This interpretation also accords with the fact that the extended induced index $m^*$, which captures second-order activity reachable through all immediate neighbors of the focal node, outperforms both the raw count $k_2$ and the original induced index $m$ in the regression models. In this sense, $m^*$ may be viewed as a useful empirical proxy for local second-order visibility, rather than as direct evidence of a specific ranking rule.

The stepwise pattern observed for the original induced index $m$, with sharp increases around $m=1$ and $m=3$, may reflect a related mechanism: some forms of second-order activity may become influential only after crossing a local reinforcement threshold. However, these threshold values were identified from the same dataset and should therefore be regarded as empirically selected. Their generality should be tested on independent data, ideally from different time periods or platforms.

\begin{figure}[t!]
	\centering
	\includegraphics[width=0.9\linewidth]{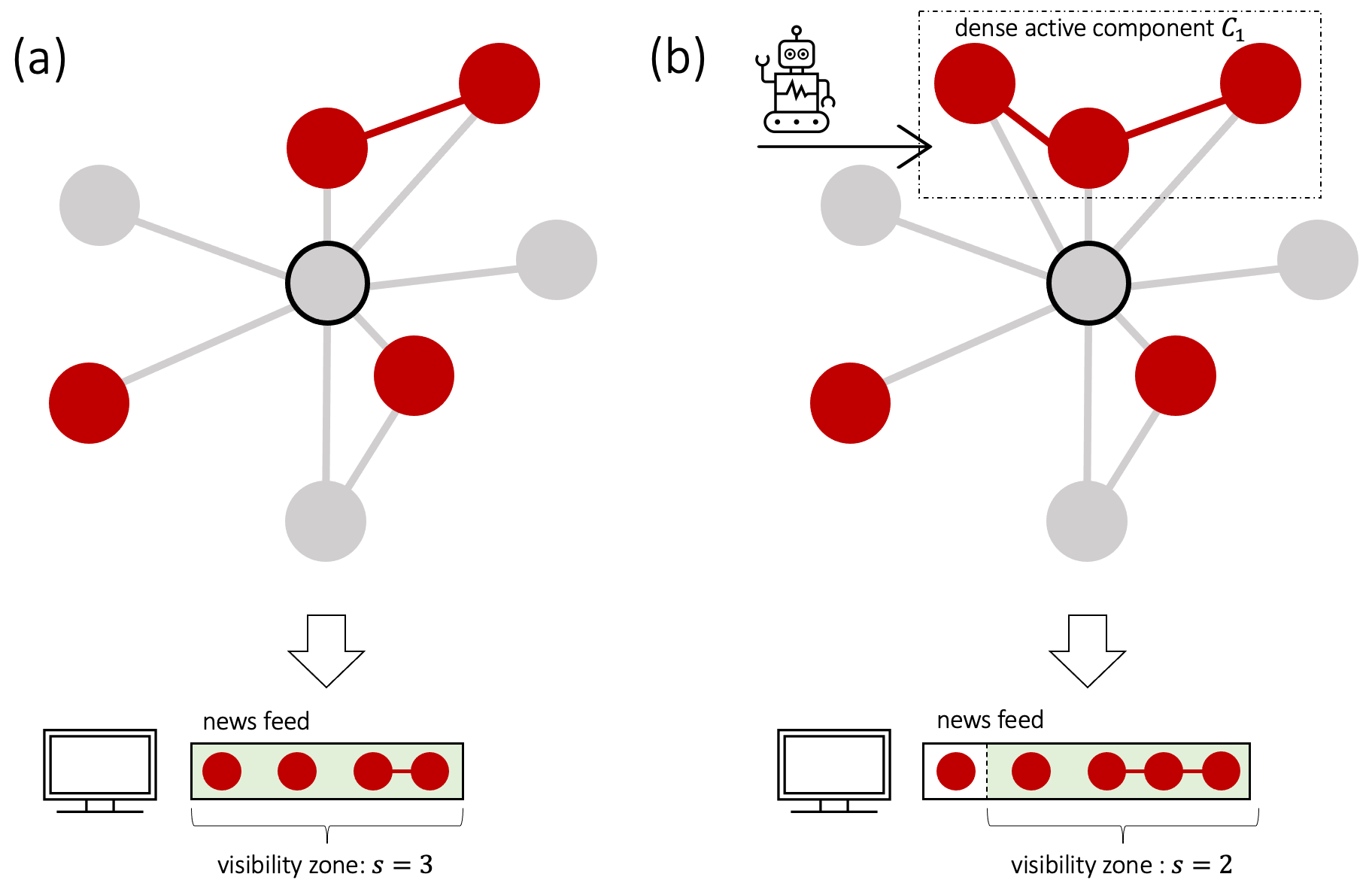}
	\caption{Illustration of the conjectured effect of increasing $k_1$ by one (active friends highlighted in maroon) while holding $s$ and $m^*$ fixed. The additional active friend increases the internal density of component $C_1$ (panel (b)). The platform's ranking algorithm may then focus the focal user's (black outline) attention on signals from this component. As a result, a user who can process only four items in the news feed observes only two connected components ($s=2$), compared with panel (a), where the measured and perceived structural diversity coincide ($s=3$).
    }\label{Fig:Explanation}
\end{figure}

The negative conditional coefficient of $k_1$ may also be interpreted through the lens of platform-mediated exposure. For fixed values of $s$ and $m^*$, an increase in $k_1$ suggests that the additional active friends are concentrated within an existing component of the active neighborhood, say community $C_1$; otherwise, the structural diversity index $s$ would tend to increase as well. Similarly, these additional active friends are unlikely to introduce new second-order activity, since this would tend to increase $m^*$. In effect, a higher value of $k_1$ at constant $s$ and $m^*$ may indicate that the focal node is more deeply embedded in a single local community. We conjecture that this deeper embedding may reduce the effective diversity of signals reaching the user for two related reasons. First, users have limited attention and cannot process the entire news feed. Second, the platform's ranking algorithms may prioritize content from the community in which the user is most active, further narrowing the user's effective exposure. As a result, signals from other components of the active neighborhood may go unnoticed, reducing the actual structural diversity experienced by the user below the value measured in the network data (see Fig.~\ref{Fig:Explanation}).
The instability of the coefficient of $\rho_1$ across specifications is consistent with this interpretation, since $\rho_1$ captures another aspect of tie concentration within the active $d=1$ neighborhood. Indeed, the scenario described above implicitly entails an increase in $\rho_1$: when $s$ and $m^*$ are held fixed, adding active friends within an existing component raises the density of ties among them, making $k_1$ and $\rho_1$ strongly correlated.

\color{black}

\subsection{Limitations}

Our study is not without limitations. First, as in most observational studies of social networks, we cannot fully control for network confounders such as homophily, common external exposure, and third-party influence~\cite{tang2025empirical}. We partially address this issue by performing robustness checks for nodes with different levels of local clustering.

Second, our interpretation in terms of platform-mediated exposure remains conjectural, because we do not observe users' actual news feeds, VKontakte's internal ranking architecture, or time-varying ranking scores. As a result, the inferred $d=2$ effects should be interpreted as predictive associations consistent with indirect exposure, rather than as direct evidence of algorithmic promotion.

Third, our analysis focuses on sufficiently popular posts, excluding posts with few likes or reposts. This selection may introduce bias, since popular posts may receive different treatment from the platform's ranking mechanisms. Relatedly, popularity-mediated visibility cannot be ruled out as an alternative explanation: posts that receive more likes may become more visible independently of the network distance between the focal user and the active set. That said, if popularity-driven visibility were the dominant mechanism, one might expect the raw count $k_2$, which is closely related to overall post reach, to be more informative than $m^*$, which captures a more structured aspect of local second-order activity. The stronger performance of $m^*$ suggests that the observed association is not reducible to raw popularity alone.

Fourth, our data are drawn from a single region (Tomsk Oblast), a single content type (local news posts), and a single platform (VKontakte). The extent to which these findings generalize to other platforms, content types, or cultural contexts remains an open question. We also note that McFadden's pseudo-$R^2$ values remain modest across all specifications, reflecting both the rarity of the outcome and the absence of richer individual-level covariates beyond gender and degree.

Fifth, we could not reconstruct repost cascades at the user level. Since each repost is also marked as a like on the parent post's page, reposts were treated as ordinary likes in the main analysis. As a robustness check, we also estimated the models on posts with no reposts, which constitute only $11\%$ of all observations. In this restricted subsample, we found no significant association with $d=1$ activity but retained a positive association between $d=2$ activity and the probability of liking.

\color{black}

\subsection*{Future Works}

In future work, we plan to extend the theoretical framework of~\cite{xie2022indirect} by developing percolation models conditioned on the extended induced index $m^*$ and validating the present findings on data from other platforms, content types, and cultural settings. Another direction for future work is to connect the empirical $d=2$ effects observed here with recent opinion-dynamics models incorporating second-neighbor or higher-order interactions (e.g.,~\cite{raineri2025}). Such models may provide a theoretical framework for understanding when indirect exposure reinforces consensus, clustering, or polarization, and how these outcomes depend on the structure of the underlying social network.

Another direction for future research concerns coordinated or inauthentic behavior in information cascades~\cite{di2025post,di2026patterns}. Our current analysis treats each post as an independent cascade and does not model the possibility that the same accounts participate systematically in multiple cascades, thereby creating dependence across diffusion processes. Moreover, we restrict attention to users in the giant connected component of the sampled Tomsk Oblast network. This choice reduces the influence of peripheral or anomalous accounts, but it also excludes a substantial share of activity: approximately $46\%$ of all likes were produced by accounts outside the analyzed component, even though this component contains $93.6\%$ of the sampled users. This disproportion suggests that excluded accounts may play a non-negligible role in the diffusion of local news posts. Future work should therefore examine whether such accounts correspond to ordinary users outside the sampled social system, organizational or media accounts, or coordinated/inauthentic activity, and how their participation affects the structure of information cascades.

\section{Data availability}

All of the data and codes can be obtained upon a reasonable request.

\section{Funding}

The research was partially supported by the TSU Development Program (Priority-2030).





\bibliography{SNA_v2}{}


\begin{thebibliography}{41}
\ifx \bisbn   \undefined \def \bisbn  #1{ISBN #1}\fi
\ifx \binits  \undefined \def \binits#1{#1}\fi
\ifx \bauthor  \undefined \def \bauthor#1{#1}\fi
\ifx \batitle  \undefined \def \batitle#1{#1}\fi
\ifx \bjtitle  \undefined \def \bjtitle#1{#1}\fi
\ifx \bvolume  \undefined \def \bvolume#1{\textbf{#1}}\fi
\ifx \byear  \undefined \def \byear#1{#1}\fi
\ifx \bissue  \undefined \def \bissue#1{#1}\fi
\ifx \bfpage  \undefined \def \bfpage#1{#1}\fi
\ifx \blpage  \undefined \def \blpage #1{#1}\fi
\ifx \burl  \undefined \def \burl#1{\textsf{#1}}\fi
\ifx \doiurl  \undefined \def \doiurl#1{\url{https://doi.org/#1}}\fi
\ifx \betal  \undefined \def \betal{\textit{et al.}}\fi
\ifx \binstitute  \undefined \def \binstitute#1{#1}\fi
\ifx \binstitutionaled  \undefined \def \binstitutionaled#1{#1}\fi
\ifx \bctitle  \undefined \def \bctitle#1{#1}\fi
\ifx \beditor  \undefined \def \beditor#1{#1}\fi
\ifx \bpublisher  \undefined \def \bpublisher#1{#1}\fi
\ifx \bbtitle  \undefined \def \bbtitle#1{#1}\fi
\ifx \bedition  \undefined \def \bedition#1{#1}\fi
\ifx \bseriesno  \undefined \def \bseriesno#1{#1}\fi
\ifx \blocation  \undefined \def \blocation#1{#1}\fi
\ifx \bsertitle  \undefined \def \bsertitle#1{#1}\fi
\ifx \bsnm \undefined \def \bsnm#1{#1}\fi
\ifx \bsuffix \undefined \def \bsuffix#1{#1}\fi
\ifx \bparticle \undefined \def \bparticle#1{#1}\fi
\ifx \barticle \undefined \def \barticle#1{#1}\fi
\bibcommenthead
\ifx \bconfdate \undefined \def \bconfdate #1{#1}\fi
\ifx \botherref \undefined \def \botherref #1{#1}\fi
\ifx \url \undefined \def \url#1{\textsf{#1}}\fi
\ifx \bchapter \undefined \def \bchapter#1{#1}\fi
\ifx \bbook \undefined \def \bbook#1{#1}\fi
\ifx \bcomment \undefined \def \bcomment#1{#1}\fi
\ifx \oauthor \undefined \def \oauthor#1{#1}\fi
\ifx \citeauthoryear \undefined \def \citeauthoryear#1{#1}\fi
\ifx \endbibitem  \undefined \def \endbibitem {}\fi
\ifx \bconflocation  \undefined \def \bconflocation#1{#1}\fi
\ifx \arxivurl  \undefined \def \arxivurl#1{\textsf{#1}}\fi
\csname PreBibitemsHook\endcsname

\bibitem[\protect\citeauthoryear{Centola and Macy}{2007}]{centola2007complex}
\begin{barticle}
\bauthor{\bsnm{Centola}, \binits{D.}},
\bauthor{\bsnm{Macy}, \binits{M.}}:
\batitle{Complex contagions and the weakness of long ties}.
\bjtitle{American Journal of Sociology}
\bvolume{113}(\bissue{3}),
\bfpage{702}--\blpage{734}
(\byear{2007})
\end{barticle}
\endbibitem

\bibitem[\protect\citeauthoryear{Kramer et~al.}{2014}]{kramer2014}
\begin{barticle}
\bauthor{\bsnm{Kramer}, \binits{A.D.I.}},
\bauthor{\bsnm{Guillory}, \binits{J.E.}},
\bauthor{\bsnm{Hancock}, \binits{J.T.}}:
\batitle{Experimental evidence of massive-scale emotional contagion through
  social networks}.
\bjtitle{Proceedings of the National Academy of Sciences}
\bvolume{111}(\bissue{24}),
\bfpage{8788}--\blpage{8790}
(\byear{2014})
\end{barticle}
\endbibitem

\bibitem[\protect\citeauthoryear{Fang et~al.}{2024}]{fang2024}
\begin{barticle}
\bauthor{\bsnm{Fang}, \binits{F.}},
\bauthor{\bsnm{Ma}, \binits{J.}},
\bauthor{\bsnm{Ma}, \binits{Y.-J.}},
\bauthor{\bsnm{Boccaletti}, \binits{S.}}:
\batitle{Social contagion on higher-order networks: The effect of relationship
  strengths}.
\bjtitle{Chaos, Solitons \& Fractals}
\bvolume{186},
\bfpage{115149}
(\byear{2024})
\end{barticle}
\endbibitem

\bibitem[\protect\citeauthoryear{Iacopini
  et~al.}{2019}]{iacopini2019simplicial}
\begin{barticle}
\bauthor{\bsnm{Iacopini}, \binits{I.}},
\bauthor{\bsnm{Petri}, \binits{G.}},
\bauthor{\bsnm{Barrat}, \binits{A.}},
\bauthor{\bsnm{Latora}, \binits{V.}}:
\batitle{Simplicial models of social contagion}.
\bjtitle{Nature Communications}
\bvolume{10},
\bfpage{2485}
(\byear{2019})
\end{barticle}
\endbibitem

\bibitem[\protect\citeauthoryear{Guess et~al.}{2023}]{guess2023}
\begin{barticle}
\bauthor{\bsnm{Guess}, \binits{A.M.}},
\bauthor{\bsnm{Malhotra}, \binits{N.}},
\bauthor{\bsnm{Pan}, \binits{J.}},
\bauthor{\bsnm{Barberá}, \binits{P.}},
\bauthor{\bsnm{Allcott}, \binits{H.}},
\bauthor{\bsnm{Brown}, \binits{T.}},
\bauthor{\bsnm{Crespo-Tenorio}, \binits{A.}},
\bauthor{\bsnm{Dimmery}, \binits{D.}},
\bauthor{\bsnm{Freelon}, \binits{D.}},
\bauthor{\bsnm{Gentzkow}, \binits{M.}},
\bauthor{\bsnm{González-Bailón}, \binits{S.}},
\bauthor{\bsnm{Kennedy}, \binits{E.}},
\bauthor{\bsnm{Kim}, \binits{Y.M.}},
\bauthor{\bsnm{Lazer}, \binits{D.}},
\bauthor{\bsnm{Moehler}, \binits{D.}},
\bauthor{\bsnm{Nyhan}, \binits{B.}},
\bauthor{\bsnm{Rivera}, \binits{C.V.}},
\bauthor{\bsnm{Settle}, \binits{J.}},
\bauthor{\bsnm{Thomas}, \binits{D.R.}},
\bauthor{\bsnm{Thorson}, \binits{E.}},
\bauthor{\bsnm{Tromble}, \binits{R.}},
\bauthor{\bsnm{Wilkins}, \binits{A.}},
\bauthor{\bsnm{Wojcieszak}, \binits{M.}},
\bauthor{\bsnm{Xiong}, \binits{B.}},
\bauthor{\bsnm{Jonge}, \binits{C.K.}},
\bauthor{\bsnm{Franco}, \binits{A.}},
\bauthor{\bsnm{Mason}, \binits{W.}},
\bauthor{\bsnm{Stroud}, \binits{N.J.}},
\bauthor{\bsnm{Tucker}, \binits{J.A.}}:
\batitle{How do social media feed algorithms affect attitudes and behavior in
  an election campaign?}
\bjtitle{Science}
\bvolume{381}(\bissue{6656}),
\bfpage{398}--\blpage{404}
(\byear{2023})
\end{barticle}
\endbibitem

\bibitem[\protect\citeauthoryear{Hodas and Lerman}{2014}]{hodas2014simple}
\begin{barticle}
\bauthor{\bsnm{Hodas}, \binits{N.O.}},
\bauthor{\bsnm{Lerman}, \binits{K.}}:
\batitle{The simple rules of social contagion}.
\bjtitle{Scientific Reports}
\bvolume{4},
\bfpage{4343}
(\byear{2014})
\end{barticle}
\endbibitem

\bibitem[\protect\citeauthoryear{Bakshy et~al.}{2015}]{bakshy2015exposure}
\begin{barticle}
\bauthor{\bsnm{Bakshy}, \binits{E.}},
\bauthor{\bsnm{Messing}, \binits{S.}},
\bauthor{\bsnm{Adamic}, \binits{L.A.}}:
\batitle{Exposure to ideologically diverse news and opinion on {F}acebook}.
\bjtitle{Science}
\bvolume{348}(\bissue{6239}),
\bfpage{1130}--\blpage{1132}
(\byear{2015})
\end{barticle}
\endbibitem

\bibitem[\protect\citeauthoryear{Christakis and
  Fowler}{2007}]{christakis2007spread}
\begin{barticle}
\bauthor{\bsnm{Christakis}, \binits{N.A.}},
\bauthor{\bsnm{Fowler}, \binits{J.H.}}:
\batitle{The spread of obesity in a large social network over 32 years}.
\bjtitle{New England Journal of Medicine}
\bvolume{357}(\bissue{4}),
\bfpage{370}--\blpage{379}
(\byear{2007})
\doiurl{10.1056/NEJMsa066082}
\end{barticle}
\endbibitem

\bibitem[\protect\citeauthoryear{Christakis and
  Fowler}{2008}]{christakis2008collective}
\begin{barticle}
\bauthor{\bsnm{Christakis}, \binits{N.A.}},
\bauthor{\bsnm{Fowler}, \binits{J.H.}}:
\batitle{The collective dynamics of smoking in a large social network}.
\bjtitle{New England Journal of Medicine}
\bvolume{358}(\bissue{21}),
\bfpage{2249}--\blpage{2258}
(\byear{2008})
\doiurl{10.1056/NEJMsa0706154}
\end{barticle}
\endbibitem

\bibitem[\protect\citeauthoryear{Xie et~al.}{2022}]{xie2022indirect}
\begin{barticle}
\bauthor{\bsnm{Xie}, \binits{J.}},
\bauthor{\bsnm{Wang}, \binits{X.}},
\bauthor{\bsnm{Feng}, \binits{L.}},
\bauthor{\bsnm{Zhao}, \binits{J.-H.}},
\bauthor{\bsnm{Liu}, \binits{W.}},
\bauthor{\bsnm{Moreno}, \binits{Y.}},
\bauthor{\bsnm{Hu}, \binits{Y.}}:
\batitle{Indirect influence in social networks as an induced percolation
  phenomenon}.
\bjtitle{Proceedings of the National Academy of Sciences}
\bvolume{119}(\bissue{9}),
\bfpage{2100151119}
(\byear{2022})
\end{barticle}
\endbibitem

\bibitem[\protect\citeauthoryear{Zhang et~al.}{2018}]{zhang2018}
\begin{barticle}
\bauthor{\bsnm{Zhang}, \binits{B.}},
\bauthor{\bsnm{Pavlou}, \binits{P.A.}},
\bauthor{\bsnm{Krishnan}, \binits{R.}}:
\batitle{On direct vs. indirect peer influence in large social networks}.
\bjtitle{Information Systems Research}
\bvolume{29}(\bissue{2}),
\bfpage{292}--\blpage{314}
(\byear{2018})
\end{barticle}
\endbibitem

\bibitem[\protect\citeauthoryear{Bakshy et~al.}{2012}]{Bakshy2012}
\begin{bchapter}
\bauthor{\bsnm{Bakshy}, \binits{E.}},
\bauthor{\bsnm{Rosenn}, \binits{I.}},
\bauthor{\bsnm{Marlow}, \binits{C.}},
\bauthor{\bsnm{Adamic}, \binits{L.}}:
\bctitle{The role of social networks in information diffusion}.
In: \bbtitle{Proceedings of the 21st International Conference on World Wide
  Web},
pp. \bfpage{519}--\blpage{528}
(\byear{2012})
\end{bchapter}
\endbibitem

\bibitem[\protect\citeauthoryear{Ugander et~al.}{2012}]{ugander2012structural}
\begin{barticle}
\bauthor{\bsnm{Ugander}, \binits{J.}},
\bauthor{\bsnm{Backstrom}, \binits{L.}},
\bauthor{\bsnm{Marlow}, \binits{C.}},
\bauthor{\bsnm{Kleinberg}, \binits{J.}}:
\batitle{Structural diversity in social contagion}.
\bjtitle{Proceedings of the National Academy of Sciences}
\bvolume{109}(\bissue{16}),
\bfpage{5962}--\blpage{5966}
(\byear{2012})
\end{barticle}
\endbibitem

\bibitem[\protect\citeauthoryear{Aral et~al.}{2009}]{aral2009pnas}
\begin{barticle}
\bauthor{\bsnm{Aral}, \binits{S.}},
\bauthor{\bsnm{Muchnik}, \binits{L.}},
\bauthor{\bsnm{Sundararajan}, \binits{A.}}:
\batitle{Distinguishing influence-based contagion from homophily-driven
  diffusion in dynamic networks}.
\bjtitle{Proceedings of the National Academy of Sciences}
\bvolume{106}(\bissue{51}),
\bfpage{21544}--\blpage{21549}
(\byear{2009})
\end{barticle}
\endbibitem

\bibitem[\protect\citeauthoryear{Backstrom et~al.}{2006}]{backstrom2006group}
\begin{bchapter}
\bauthor{\bsnm{Backstrom}, \binits{L.}},
\bauthor{\bsnm{Huttenlocher}, \binits{D.}},
\bauthor{\bsnm{Kleinberg}, \binits{J.}},
\bauthor{\bsnm{Lan}, \binits{X.}}:
\bctitle{Group formation in large social networks: membership, growth, and
  evolution}.
In: \bbtitle{Proceedings of the 12th ACM SIGKDD International Conference on
  Knowledge Discovery and Data Mining},
pp. \bfpage{44}--\blpage{54}
(\byear{2006})
\end{bchapter}
\endbibitem

\bibitem[\protect\citeauthoryear{Kairam et~al.}{2012}]{kairam2012life}
\begin{bchapter}
\bauthor{\bsnm{Kairam}, \binits{S.R.}},
\bauthor{\bsnm{Wang}, \binits{D.J.}},
\bauthor{\bsnm{Leskovec}, \binits{J.}}:
\bctitle{The life and death of online groups: Predicting group growth and
  longevity}.
In: \bbtitle{Proceedings of the Fifth ACM International Conference on Web
  Search and Data Mining},
pp. \bfpage{673}--\blpage{682}
(\byear{2012})
\end{bchapter}
\endbibitem

\bibitem[\protect\citeauthoryear{Centola}{2010}]{centola2010spread}
\begin{barticle}
\bauthor{\bsnm{Centola}, \binits{D.}}:
\batitle{The spread of behavior in an online social network experiment}.
\bjtitle{Science}
\bvolume{329}(\bissue{5996}),
\bfpage{1194}--\blpage{1197}
(\byear{2010})
\end{barticle}
\endbibitem

\bibitem[\protect\citeauthoryear{Aral and Nicolaides}{2017}]{aral2017exercise}
\begin{barticle}
\bauthor{\bsnm{Aral}, \binits{S.}},
\bauthor{\bsnm{Nicolaides}, \binits{C.}}:
\batitle{Exercise contagion in a global social network}.
\bjtitle{Nature Communications}
\bvolume{8}(\bissue{1}),
\bfpage{14753}
(\byear{2017})
\end{barticle}
\endbibitem

\bibitem[\protect\citeauthoryear{M{\o}nsted et~al.}{2017}]{monsted2017evidence}
\begin{barticle}
\bauthor{\bsnm{M{\o}nsted}, \binits{B.}},
\bauthor{\bsnm{Sapie{\.z}y{\'n}ski}, \binits{P.}},
\bauthor{\bsnm{Ferrara}, \binits{E.}},
\bauthor{\bsnm{Lehmann}, \binits{S.}}:
\batitle{Evidence of complex contagion of information in social media: An
  experiment using {T}witter bots}.
\bjtitle{PloS one}
\bvolume{12}(\bissue{9}),
\bfpage{0184148}
(\byear{2017})
\end{barticle}
\endbibitem

\bibitem[\protect\citeauthoryear{Lee et~al.}{2025}]{lee2025complex}
\begin{barticle}
\bauthor{\bsnm{Lee}, \binits{J.}},
\bauthor{\bsnm{Lazer}, \binits{D.}},
\bauthor{\bsnm{Riedl}, \binits{C.}}:
\batitle{Complex contagion in social networks: Causal evidence from a
  country-scale field experiment}.
\bjtitle{Sociological Science}
\bvolume{12},
\bfpage{685}--\blpage{714}
(\byear{2025})
\end{barticle}
\endbibitem

\bibitem[\protect\citeauthoryear{Romero et~al.}{2011}]{romero2011differences}
\begin{bchapter}
\bauthor{\bsnm{Romero}, \binits{D.M.}},
\bauthor{\bsnm{Meeder}, \binits{B.}},
\bauthor{\bsnm{Kleinberg}, \binits{J.}}:
\bctitle{Differences in the mechanics of information diffusion across topics:
  idioms, political hashtags, and complex contagion on {T}witter}.
In: \bbtitle{Proceedings of the 20th International Conference on World Wide
  Web},
pp. \bfpage{695}--\blpage{704}
(\byear{2011})
\end{bchapter}
\endbibitem

\bibitem[\protect\citeauthoryear{Wang et~al.}{2019}]{wang2019anomalous}
\begin{barticle}
\bauthor{\bsnm{Wang}, \binits{X.}},
\bauthor{\bsnm{Lan}, \binits{Y.}},
\bauthor{\bsnm{Xiao}, \binits{J.}}:
\batitle{Anomalous structure and dynamics in news diffusion among heterogeneous
  individuals}.
\bjtitle{Nature Human Behaviour}
\bvolume{3}(\bissue{7}),
\bfpage{709}--\blpage{718}
(\byear{2019})
\end{barticle}
\endbibitem

\bibitem[\protect\citeauthoryear{Centola}{2019}]{centola2019influential}
\begin{barticle}
\bauthor{\bsnm{Centola}, \binits{D.}}:
\batitle{Influential networks}.
\bjtitle{Nature Human Behaviour}
\bvolume{3}(\bissue{7}),
\bfpage{664}--\blpage{665}
(\byear{2019})
\end{barticle}
\endbibitem

\bibitem[\protect\citeauthoryear{Guilbeault and
  Centola}{2021}]{guilbeault2021topological}
\begin{barticle}
\bauthor{\bsnm{Guilbeault}, \binits{D.}},
\bauthor{\bsnm{Centola}, \binits{D.}}:
\batitle{Topological measures for identifying and predicting the spread of
  complex contagions}.
\bjtitle{Nature Communications}
\bvolume{12}(\bissue{1}),
\bfpage{4430}
(\byear{2021})
\end{barticle}
\endbibitem

\bibitem[\protect\citeauthoryear{Airoldi and
  Christakis}{2024}]{airoldi2024induction}
\begin{barticle}
\bauthor{\bsnm{Airoldi}, \binits{E.M.}},
\bauthor{\bsnm{Christakis}, \binits{N.A.}}:
\batitle{Induction of social contagion for diverse outcomes in structured
  experiments in isolated villages}.
\bjtitle{Science}
\bvolume{384}(\bissue{6695}),
\bfpage{5147}
(\byear{2024})
\end{barticle}
\endbibitem

\bibitem[\protect\citeauthoryear{Li et~al.}{2024}]{li2024social}
\begin{barticle}
\bauthor{\bsnm{Li}, \binits{T.}},
\bauthor{\bsnm{Wu}, \binits{Y.}},
\bauthor{\bsnm{Ding}, \binits{Q.}},
\bauthor{\bsnm{Xie}, \binits{Y.}},
\bauthor{\bsnm{Yu}, \binits{D.}},
\bauthor{\bsnm{Yang}, \binits{L.}},
\bauthor{\bsnm{Jia}, \binits{Y.}}:
\batitle{Social contagion in high-order network with mutation}.
\bjtitle{Chaos, Solitons \& Fractals}
\bvolume{180},
\bfpage{114583}
(\byear{2024})
\end{barticle}
\endbibitem

\bibitem[\protect\citeauthoryear{de~Miguel-Arribas
  et~al.}{2025}]{de2025interplay}
\begin{barticle}
\bauthor{\bsnm{Miguel-Arribas}, \binits{A.}},
\bauthor{\bsnm{Aleta}, \binits{A.}},
\bauthor{\bsnm{Moreno}, \binits{Y.}}:
\batitle{Interplay of epidemic spreading and vaccine uptake under complex
  social contagion}.
\bjtitle{Physical Review E}
\bvolume{112}(\bissue{1}),
\bfpage{014308}
(\byear{2025})
\end{barticle}
\endbibitem

\bibitem[\protect\citeauthoryear{Komander et~al.}{}]{komander2025modeling}
\begin{botherref}
\oauthor{\bsnm{Komander}, \binits{B.}},
\oauthor{\bsnm{Cerquides}, \binits{J.}},
\oauthor{\bsnm{Chan}, \binits{J.}},
\oauthor{\bsnm{Alavi}, \binits{A.}}:
Modeling link recommendations as a network growth mechanism and their impact on
  social contagion.
In: ICLR 2025 Workshop on Human-AI Coevolution
\end{botherref}
\endbibitem

\bibitem[\protect\citeauthoryear{Fowler and
  Christakis}{2010}]{fowler2010cooperative}
\begin{barticle}
\bauthor{\bsnm{Fowler}, \binits{J.H.}},
\bauthor{\bsnm{Christakis}, \binits{N.A.}}:
\batitle{Cooperative behavior cascades in human social networks}.
\bjtitle{Proceedings of the National Academy of Sciences}
\bvolume{107}(\bissue{12}),
\bfpage{5334}--\blpage{5338}
(\byear{2010})
\end{barticle}
\endbibitem

\bibitem[\protect\citeauthoryear{Christakis and
  Fowler}{2013}]{christakis2013social}
\begin{barticle}
\bauthor{\bsnm{Christakis}, \binits{N.A.}},
\bauthor{\bsnm{Fowler}, \binits{J.H.}}:
\batitle{Social contagion theory: examining dynamic social networks and human
  behavior}.
\bjtitle{Statistics in Medicine}
\bvolume{32}(\bissue{4}),
\bfpage{556}--\blpage{577}
(\byear{2013})
\end{barticle}
\endbibitem

\bibitem[\protect\citeauthoryear{Guimar{\~a}es~Jr
  et~al.}{2017}]{guimaraes2017indirect}
\begin{barticle}
\bauthor{\bsnm{Guimar{\~a}es~Jr}, \binits{P.R.}},
\bauthor{\bsnm{Pires}, \binits{M.M.}},
\bauthor{\bsnm{Jordano}, \binits{P.}},
\bauthor{\bsnm{Bascompte}, \binits{J.}},
\bauthor{\bsnm{Thompson}, \binits{J.N.}}:
\batitle{Indirect effects drive coevolution in mutualistic networks}.
\bjtitle{Nature}
\bvolume{550}(\bissue{7677}),
\bfpage{511}--\blpage{514}
(\byear{2017})
\end{barticle}
\endbibitem

\bibitem[\protect\citeauthoryear{Miranda et~al.}{2024}]{Miranda2024}
\begin{barticle}
\bauthor{\bsnm{Miranda}, \binits{M.}},
\bauthor{\bsnm{Pereda}, \binits{M.}},
\bauthor{\bsnm{Sánchez}, \binits{A.}},
\bauthor{\bsnm{Estrada}, \binits{E.}}:
\batitle{Indirect social influence and diffusion of innovations: An
  experimental approach}.
\bjtitle{PNAS Nexus}
\bvolume{3}(\bissue{10}),
\bfpage{409}
(\byear{2024})
\end{barticle}
\endbibitem

\bibitem[\protect\citeauthoryear{VanderWeele}{2013}]{vanderWeele2013}
\begin{barticle}
\bauthor{\bsnm{VanderWeele}, \binits{T.J.}}:
\batitle{Inference for influence over multiple degrees of separation on a
  social network}.
\bjtitle{Statistics in Medicine}
\bvolume{32}(\bissue{4}),
\bfpage{591}--\blpage{596}
(\byear{2013})
\end{barticle}
\endbibitem

\bibitem[\protect\citeauthoryear{Gonz{\'a}lez-Bail{\'o}n and
  De~Domenico}{2021}]{gonzalez2021bots}
\begin{barticle}
\bauthor{\bsnm{Gonz{\'a}lez-Bail{\'o}n}, \binits{S.}},
\bauthor{\bsnm{De~Domenico}, \binits{M.}}:
\batitle{Bots are less central than verified accounts during contentious
  political events}.
\bjtitle{Proceedings of the National Academy of Sciences}
\bvolume{118}(\bissue{11}),
\bfpage{2013443118}
(\byear{2021})
\end{barticle}
\endbibitem

\bibitem[\protect\citeauthoryear{Watts and
  Strogatz}{1998}]{watts1998collective}
\begin{barticle}
\bauthor{\bsnm{Watts}, \binits{D.J.}},
\bauthor{\bsnm{Strogatz}, \binits{S.H.}}:
\batitle{Collective dynamics of 'small-world' networks}.
\bjtitle{Nature}
\bvolume{393},
\bfpage{440}--\blpage{442}
(\byear{1998})
\end{barticle}
\endbibitem

\bibitem[\protect\citeauthoryear{Long and Freese}{2021}]{long2021regression}
\begin{bbook}
\bauthor{\bsnm{Long}, \binits{J.S.}},
\bauthor{\bsnm{Freese}, \binits{J.}}:
\bbtitle{Regression Models for Categorical Dependent Variables Using Stata}.
\bpublisher{Stata Press},
\blocation{College Station, TX}
(\byear{2021})
\end{bbook}
\endbibitem

\bibitem[\protect\citeauthoryear{Tang et~al.}{2025}]{tang2025empirical}
\begin{botherref}
\oauthor{\bsnm{Tang}, \binits{T.}},
\oauthor{\bsnm{Snijders}, \binits{T.A.}},
\oauthor{\bsnm{Flache}, \binits{A.}}:
An empirical and simulation investigation of bounded confidence and negative
  influence in opinion dynamics using stochastic actor-oriented model.
JASSS
\textbf{28}(1)
(2025)
\end{botherref}
\endbibitem

\bibitem[\protect\citeauthoryear{Carullo et~al.}{2015}]{carullo2015triadic}
\begin{barticle}
\bauthor{\bsnm{Carullo}, \binits{G.}},
\bauthor{\bsnm{Castiglione}, \binits{A.}},
\bauthor{\bsnm{De~Santis}, \binits{A.}},
\bauthor{\bsnm{Palmieri}, \binits{F.}}:
\batitle{A triadic closure and homophily-based recommendation system for online
  social networks}.
\bjtitle{World Wide Web}
\bvolume{18}(\bissue{6}),
\bfpage{1579}--\blpage{1601}
(\byear{2015})
\end{barticle}
\endbibitem

\bibitem[\protect\citeauthoryear{Raineri et~al.}{2025}]{raineri2025}
\begin{barticle}
\bauthor{\bsnm{Raineri}, \binits{R.}},
\bauthor{\bsnm{Zino}, \binits{L.}},
\bauthor{\bsnm{Proskurnikov}, \binits{A.}}:
\batitle{{FJ-MM}: {F}riedkin–{J}ohnsen opinion dynamics model with memory and
  higher-order neighbors}.
\bjtitle{European Journal of Control}
\bvolume{86},
\bfpage{101306}
(\byear{2025})
\end{barticle}
\endbibitem

\bibitem[\protect\citeauthoryear{Di~Marco et~al.}{2025}]{di2025post}
\begin{barticle}
\bauthor{\bsnm{Di~Marco}, \binits{N.}},
\bauthor{\bsnm{Brunetti}, \binits{S.}},
\bauthor{\bsnm{Cinelli}, \binits{M.}},
\bauthor{\bsnm{Quattrociocchi}, \binits{W.}}:
\batitle{Post-hoc evaluation of nodes influence in information cascades: The
  case of coordinated accounts}.
\bjtitle{ACM Transactions on the Web}
\bvolume{19}(\bissue{2}),
\bfpage{1}--\blpage{19}
(\byear{2025})
\end{barticle}
\endbibitem

\bibitem[\protect\citeauthoryear{Di~Marco et~al.}{2026}]{di2026patterns}
\begin{barticle}
\bauthor{\bsnm{Di~Marco}, \binits{N.}},
\bauthor{\bsnm{Bonetti}, \binits{A.}},
\bauthor{\bsnm{Di~Martino}, \binits{E.}},
\bauthor{\bsnm{Loru}, \binits{E.}},
\bauthor{\bsnm{Nudo}, \binits{J.}},
\bauthor{\bsnm{Pandolfo}, \binits{M.E.}},
\bauthor{\bsnm{Pecile}, \binits{G.}},
\bauthor{\bsnm{Sangiorgio}, \binits{E.}},
\bauthor{\bsnm{Scalco}, \binits{I.}},
\bauthor{\bsnm{Zollo}, \binits{S.}},
\bauthor{\bsnm{Cinelli}, \binits{M.}},
\bauthor{\bsnm{Zollo}, \binits{F.}},
\bauthor{\bsnm{Quattrociocchi}, \binits{W.}}:
\batitle{Patterns, models, and challenges in online social media: a survey}.
\bjtitle{ACM Transactions on the Web}
\bvolume{20}(\bissue{2}),
\bfpage{1}--\blpage{34}
(\byear{2026})
\end{barticle}
\endbibitem

\end{thebibliography}

\begin{table}[h]
\renewcommand{\arraystretch}{3}
\tiny
\centering
\caption{Regression fit (robustness checks); coefficient estimates are reported along with standard errors (in parentheses).}
\label{tab:RegressionFit_checks}
\begin{tabular}{>{\raggedright\arraybackslash}m{0.15\linewidth}>{\centering\arraybackslash}m{0.1\linewidth}>{\centering\arraybackslash}m{0.1\linewidth}>{\centering\arraybackslash}m{0.1\linewidth}>{\centering\arraybackslash}m{0.1\linewidth}>{\centering\arraybackslash}m{0.1\linewidth}>{\centering\arraybackslash}m{0.1\linewidth}}
\hline
    \textbf{Variable}
    & \makecell[c]{$\mathcal{M}_{9}$\\[1pt]{baseline}}
    & \makecell[c]{$\mathcal{M}_{9}$\\[1pt]{low-degree}\\{nodes}}
    & \makecell[c]{$\mathcal{M}_{9}$\\[1pt]{high-degree}\\{nodes}}
    & \makecell[c]{$\mathcal{M}_{10}$\\[1pt]{baseline}}
    & \makecell[c]{$\mathcal{M}_{10}$\\[1pt]{low-degree}\\{nodes}}
    & \makecell[c]{$\mathcal{M}_{10}$\\[1pt]{high-degree}\\{nodes}} \\

    \hline

    Intercept
      & \makecell[c]{$-8.7$ \\ $(0.010)$}
      & \makecell[c]{$-9.4$ \\ $(0.035)$}
      & \makecell[c]{$-8.6$ \\ $(0.017)$}
      & \makecell[c]{$-8.5$ \\ $(0.0100)$}
      & \makecell[c]{$-9.4$ \\ $(0.035)$}
      & \makecell[c]{$-8.3$ \\ $(0.016)$} \\

    Active friends $k_1$
      & \makecell[c]{$-1.2$ \\ $(0.18)$}
      & \makecell[c]{$16$ \\ \textbf{P=1.000} \\ $(8.8\times 10^6)$}
      & \makecell[c]{$-1.1$ \\ $(0.18)$}
      & \makecell[c]{$-0.25$ \\ \textbf{P=0.081} \\ $(0.14)$}
      & \makecell[c]{$-5.0$ \\ \textbf{P=0.943} \\ $(70)$}
      & \makecell[c]{$-0.18$ \\ \textbf{P=0.191} \\ $(0.14)$} \\

    Density of active $d=1$ network $\rho_1$
      & \makecell[c]{$0.38$ \\ \textbf{P=0.172} \\ $(0.28)$}
      & \makecell[c]{$-200$ \\ \textbf{P=1.000} \\ $(1.5 \times 10^{12})$}
      & \makecell[c]{$0.36$ \\ \textbf{P=0.204} \\ $(0.28)$}
      & \makecell[c]{$-1.7$ \\ $(0.25)$}
      & \makecell[c]{$-0.62$ \\ \textbf{P=0.993} \\ $(73)$}
      & \makecell[c]{$-1.7$ \\ $(0.26)$} \\

    Structural diversity $s$
      & \makecell[c]{$1.1$ \\ $(0.18)$}
      & \makecell[c]{$-15$ \\ \textbf{P=1.000} \\ $(8.8\times 10^6)$}
      & \makecell[c]{$0.92$ \\ $(0.18)$}
      & \makecell[c]{$-2.1$ \\ $(0.17)$}
      & \makecell[c]{$5.2$ \\ \textbf{P=0.941} \\ $(70)$}
      & \makecell[c]{$-2.4$ \\ $(0.17)$} \\

    Extended induced index $m^*$
      & \makecell[c]{$0.083$ \\ $(0.00050)$}
      & \makecell[c]{$0.096$ \\ $(0.0017)$}
      & \makecell[c]{$0.078$ \\ $(0.00065)$}
      & ---
      & ---
      & --- \\

    Indicator $\indi{1 \leq m \leq 3}$
      & ---
      & ---
      & ---
      & \makecell[c]{$5.5$ \\ $(0.088)$}
      & \makecell[c]{$3.2$ \\ $(0.31)$}
      & \makecell[c]{$5.6$ \\ $(0.095)$} \\

    Indicator $\indi{m>3}$
      & ---
      & ---
      & ---
      & \makecell[c]{$6.3$ \\ $(0.097)$}
      & \makecell[c]{$3.0$ \\ $(0.49)$}
      & \makecell[c]{$6.5$ \\ $(0.10)$} \\

    Male gender
      & \makecell[c]{$-0.050$ \\ $(0.015)$}
      & \makecell[c]{$-0.37$ \\ $(0.034)$}
      & \makecell[c]{$0.091$ \\ $(0.021)$}
      & \makecell[c]{$-0.079$ \\ $(0.015)$}
      & \makecell[c]{$-0.39$ \\ $(0.034)$}
      & \makecell[c]{$0.067$ \\ \textbf{P=0.002} \\ $(0.021)$} \\

    Degree
      & \makecell[c]{$0.00032$ \\ $(0.000023)$}
      & \makecell[c]{$0.054$ \\ $(0.0028)$}
      & \makecell[c]{$0.00019$ \\ $(0.000028)$}
      & \makecell[c]{$0.00011$ \\ $(0.000025)$}
      & \makecell[c]{$0.062$ \\ $(0.0028)$}
      & \makecell[c]{$-0.00011$ \\ $(0.000032)$} \\

    \hline

    McFadden's pseudo-$R^2$
      & $0.0388$ & $0.0265$ & $0.0499$ & $0.0259$ & $0.0121$ & $0.0451$ \\

    BIC
      & 354,593.39 & 74,326.63 & $155{,}458.17$ & $359{,}360.60$ & $75{,}445.15$ & $156{,}259.08$ \\

    \hline
\end{tabular}
\begin{tablenotes}
         \item[] Unless an exact $P$-value is indicated, all estimates are statistically significant at $P<0.001$.
         \item[] Extremely large standard errors in the low-degree subsample reflect sparse exposure categories and should not be interpreted substantively.
\end{tablenotes}
\end{table}

\begin{table}[h]
\renewcommand{\arraystretch}{3}
\tiny
\centering
\caption{Regression fit (robustness checks, part 2); coefficient estimates are reported along with standard errors (in parentheses).}
\label{tab:RegressionFit_checks_(part_2)}
\begin{tabular}{>{\raggedright\arraybackslash}m{0.15\linewidth}>{\centering\arraybackslash}m{0.1\linewidth}>{\centering\arraybackslash}m{0.1\linewidth}>{\centering\arraybackslash}m{0.1\linewidth}>{\centering\arraybackslash}m{0.1\linewidth}>{\centering\arraybackslash}m{0.1\linewidth}>{\centering\arraybackslash}m{0.1\linewidth}}
\hline
    \textbf{Variable}
    & \makecell[c]{$\mathcal{M}_{9}$\\[1pt]{baseline}}
    & \makecell[c]{$\mathcal{M}_{9}$\\[1pt]{low-clustering}\\{nodes}}
    & \makecell[c]{$\mathcal{M}_{9}$\\[1pt]{high-clustering}\\{nodes}}
    & \makecell[c]{$\mathcal{M}_{10}$\\[1pt]{baseline}}
    & \makecell[c]{$\mathcal{M}_{10}$\\[1pt]{low-clustering}\\{nodes}}
    & \makecell[c]{$\mathcal{M}_{10}$\\[1pt]{high-clustering}\\{nodes}} \\

    \hline

    Intercept
      & \makecell[c]{$-8.7$ \\ $(0.010)$}
      & \makecell[c]{$-8.4$ \\ $(0.015)$}
      & \makecell[c]{$-9.1$ \\ $(0.022)$}
      & \makecell[c]{$-8.5$ \\ $(0.0100)$}
      & \makecell[c]{$-8.2$ \\ $(0.015)$}
      & \makecell[c]{$-8.9$ \\ $(0.021)$} \\

    Active friends $k_1$
      & \makecell[c]{$-1.2$ \\ $(0.18)$}
      & \makecell[c]{$-2.0$ \\ $(0.51)$}
      & \makecell[c]{$-0.94$ \\ $(0.22)$}
      & \makecell[c]{$-0.25$ \\ \textbf{P=0.081} \\ $(0.14)$}
      & \makecell[c]{$-0.30$ \\ \textbf{P=0.406} \\ $(0.36)$}
      & \makecell[c]{$-0.29$ \\ \textbf{P=0.133} \\ $(0.19)$} \\

    Density of active $d = 1$ network $\rho_1$
      & \makecell[c]{$0.38$ \\ \textbf{P=0.172} \\ $(0.28)$}
      & \makecell[c]{$1.8$ \\ \textbf{P=0.004} \\ $(0.62)$}
      & \makecell[c]{$-0.026$ \\ \textbf{P=0.949} \\ $(0.40)$}
      & \makecell[c]{$-1.7$ \\ $(0.25)$}
      & \makecell[c]{$-0.94$ \\ \textbf{P=0.057} \\ $(0.49)$}
      & \makecell[c]{$-1.7$ \\ $(0.38)$} \\

    Structural diversity $s$
      & \makecell[c]{$1.1$ \\ $(0.18)$}
      & \makecell[c]{$1.8$ \\ $(0.52)$}
      & \makecell[c]{$0.83$ \\ $(0.23)$}
      & \makecell[c]{$-2.1$ \\ $(0.17)$}
      & \makecell[c]{$-2.2$ \\ $(0.39)$}
      & \makecell[c]{$-1.4$ \\ $(0.25)$} \\

    Extended induced index $m^*$
      & \makecell[c]{$0.083$ \\ $(0.00050)$}
      & \makecell[c]{$0.083$ \\ $(0.00076)$}
      & \makecell[c]{$0.084$ \\ $(0.0011)$}
      & ---
      & ---
      & --- \\

    $\indi{1 \leq m \leq 3}$
      & ---
      & ---
      & ---
      & \makecell[c]{$5.5$ \\ $(0.088)$}
      & \makecell[c]{$5.9$ \\ $(0.12)$}
      & \makecell[c]{$4.1$ \\ $(0.17)$} \\

    $\indi{m>3}$
      & ---
      & ---
      & ---
      & \makecell[c]{$6.3$ \\ $(0.097)$}
      & \makecell[c]{$6.6$ \\ $(0.13)$}
      & \makecell[c]{$5.6$ \\ $(0.18)$} \\

    Degree
      & \makecell[c]{$0.00032$ \\ $(0.000023)$}
      & \makecell[c]{$0.00021$ \\ $(0.000050)$}
      & \makecell[c]{$0.00060$ \\ $(0.000038)$}
      & \makecell[c]{$0.00011$ \\ $(0.000025)$}
      & \makecell[c]{$0.00013$ \\ \textbf{P=0.003} \\ $(0.000045)$}
      & \makecell[c]{$0.00046$ \\ $(0.000043)$} \\

    Male gender
      & \makecell[c]{$-0.050$ \\ $(0.015)$}
      & \makecell[c]{$-0.19$ \\ $(0.023)$}
      & \makecell[c]{$0.061$ \\ \textbf{P=0.040} \\ $(0.030)$}
      & \makecell[c]{$-0.079$ \\ $(0.015)$}
      & \makecell[c]{$-0.22$ \\ $(0.023)$}
      & \makecell[c]{$0.027$ \\ \textbf{P=0.356} \\ $(0.030)$} \\

    \hline

    McFadden's pseudo-$R^2$
      & $0.0388$ & $0.0425$ & $0.0333$ & $0.0259$ & $0.0318$ & $0.0208$ \\
    BIC
      & $354{,}593.39$ & $146{,}357.35$ & $88{,}403.32$ & $359{,}360.60$ & $148{,}005.30$ & $89{,}562.43$ \\

    \hline
\end{tabular}
\begin{tablenotes}
     \item[] Unless an exact $P$-value is indicated, all estimates are statistically significant at $P<0.001$.
\end{tablenotes}
\end{table}

\end{document}